\documentclass[aps,a4paper,twocolumn]{revtex4-2}

\usepackage{amsmath}
\usepackage{graphicx}
\usepackage{subfigure}
\usepackage[usenames,dvipsnames]{color}
\definecolor{darkblue}{RGB}{0,0,196}
\usepackage[colorlinks=true,linkcolor=darkblue,citecolor=darkblue,urlcolor=darkblue]{hyperref}
\usepackage{setspace}
\usepackage{multirow}
\usepackage{hyperref}
\usepackage{xcolor}
\hypersetup{
  colorlinks   = true, 
  urlcolor     = blue, 
  linkcolor    = blue, 
  citecolor   = blue 
}
\DeclareUnicodeCharacter{2212}{-}
\usepackage{graphicx}
\usepackage{amsmath,bbm}
\usepackage{amssymb,bm}
\usepackage{yfonts}
\usepackage{comment}
\usepackage[normalem]{ulem}
\begin{document}

\title{Inclusive, prompt and non-prompt $\rm{J}/\psi$ identification in proton-proton collisions at the Large Hadron Collider using machine learning }

\author{Suraj Prasad}\email[]{Suraj.Prasad@cern.ch}
\author{Neelkamal Mallick}\email[]{Neelkamal.Mallick@cern.ch}
\author{Raghunath Sahoo}\email[Corresponding author: ]{Raghunath.Sahoo@cern.ch}
\affiliation{Department of Physics, Indian Institute of Technology Indore, Simrol, Indore 453552, India}

\date{\today}
\begin{abstract}

Studies related to $\rm{J}/\psi$ meson, a bound state of charm and anti-charm quarks ($c\bar{c}$), in heavy-ion collisions, provide genuine testing grounds for the theory of strong interaction, quantum chromodynamics (QCD). To better understand the underlying production mechanism, cold nuclear matter effects, and influence from the quark-gluon plasma, baseline measurements are also performed in proton-proton ($pp$) and proton-nucleus ($p$--A) collisions. The inclusive $\rm{J}/\psi$ measurement has contributions from both prompt and non-prompt productions. The prompt $\rm{J}/\psi$ is produced directly from the hadronic interactions or via feed-down from directly produced higher charmonium states, whereas non-prompt $\rm{J}/\psi$ comes from the decay of beauty hadrons.  In experiments, $\rm{J}/\psi$ is reconstructed through its electromagnetic decays to lepton pairs, in either $e^{+}+e^{-}$ or $\mu^{+}+\mu^{-}$ decay channels. In this work, for the first time, machine learning techniques are implemented to separate the prompt and non-prompt dimuon pairs from the background to obtain a better identification of the $\rm{J}/\psi$ signal for different production modes. The study has been performed in $pp$ collisions at $\sqrt{s} = 7$ and 13 TeV simulated using PYTHIA8. Machine learning models such as XGBoost and LightGBM are explored. The models could achieve up to 99\% prediction accuracy. The transverse momentum ($p_{\rm T}$)  and rapidity ($y$) differential measurements of inclusive, prompt, and non-prompt $\rm{J}/\psi$, its multiplicity dependence, and the $p_{\rm T}$ dependence of fraction of non-prompt $\rm{J}/\psi$  ($f_{\rm B}$) are shown. These results are compared to experimental findings wherever possible.

\end{abstract}

\maketitle 

\section{Introduction}
\label{sec:intro}
Over the last couple of decades, two of the world's most powerful particle accelerators, the Large Hadron Collider (LHC), CERN, and the Relativistic Heavy-Ion Collider (RHIC), Brookhaven National Laboratory, USA have studied the hot and dense state of deconfined partons, known as the quark-gluon plasma (QGP) by colliding heavy-ions at ultra-relativistic speeds. These studies are crucial to understand the physics of the early Universe, and the phase transition between the partonic and hadronic matter. Due to the nature of the strong interaction, QGP is extremely short-lived. Therefore, to study the properties of QGP, several indirect signatures are investigated. One such signature is the melting of heavy quarkonia ($q\bar{q}$) in QGP, also known as the quarkonia suppression, where the color force responsible for binding the quarks into hadrons is screened in the presence of deconfined partons~\cite{Matsui:1986dk, Digal:2001ue,ALICE:2016flj,ALICE:2013osk,ALICE:2014wnc, ALICE:2020wwx}. The production of heavy quarkonia pairs ($c\bar{c}$ and $b\bar{b}$) follow the perturbative QCD (pQCD) calculations, whereas the evolution to a bound colorless state is a nonperturbative process. Due to their high mass, heavy-quarks are produced via partonic interactions in the early stages of the collision, and experience the full evolution of QGP. Thus, they are sensitive probes to study the properties of QGP and the theory of strong interaction~\cite{Shuryak:1978ij}.

\begin{figure}[ht!]
\includegraphics[scale=0.7]{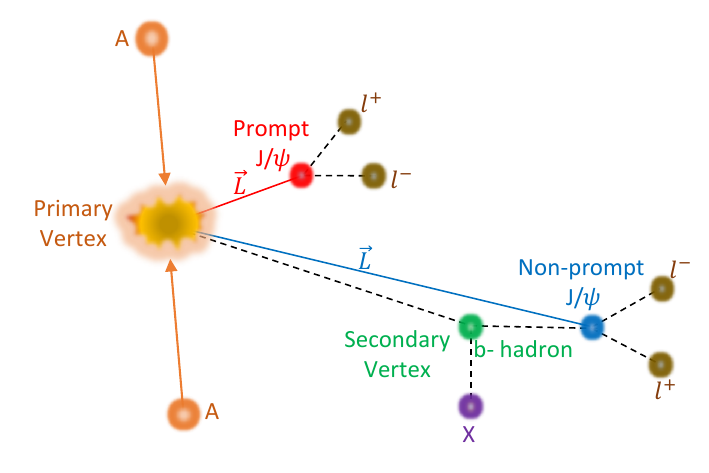}
\caption{Schematic representation of topological production of prompt and non-prompt $\rm{J}/\psi$ mesons in hadronic and nuclear collisions.}
\label{fig:jpsicartoon}
\end{figure}

$\rm{J}/\psi$ is the lightest charm vector meson, which is the bound state of a charm and an anti-charm quark ($c\bar{c}$). The studies related to $\rm{J}/\psi$ meson, in heavy-ion collisions provide genuine testing grounds for QCD~\cite{ALICE:2015nvt,ALICE:2018cqy}. To better understand the underlying production mechanism, cold nuclear matter effects, and influence from the quark-gluon plasma, baseline measurements are also performed in proton-proton ($pp$) and proton-nucleus ($p$--A) collisions~\cite{ALICE:2021dtt,ALICE:2022zig}. The inclusive $\rm{J}/\psi$ production can have contributions from three sources. The first one is the direct prompt production, in which $\rm{J}/\psi$ is produced directly from the hadronic/nuclear collisions; the second one is the indirect prompt production via feed-down from directly produced higher charmonium states (\textit{i.e.} from $\chi_c$ and $\psi(\rm 2S)$), and the third one is the non-prompt production which comes from the decay of beauty hadrons~\cite{ALICE:2012vpz,ALICE:2018szk}. Figure~\ref{fig:jpsicartoon} depicts the topological productions of $\rm{J}/\psi$, where $\vec{L}$ denotes the vector joining the $\rm{J}/\psi$ decay vertex to the primary vertex. In Fig.~\ref{fig:jpsicartoon}, it is evident that the prompt $\rm{J}/\psi$ is produced nearer to the primary vertex compared to the non-prompt $\rm{J}/\psi$, where b-hadrons fly off to a finite distance before decaying to $\rm{J}/\psi$ via weak decay. Since the rest mass of $\rm{J}/\psi$ is larger than the other decay daughters of beauty hadron, the momentum of $\rm{J}/\psi$ is closer to the decaying beauty hadron, thus non-prompt $\rm{J}/\psi$ gives a better handle to study the production of these beauty hadrons \cite{CDF:2004jtw}. Another important implication of separating the non-prompt $\rm{J}/\psi$ from prompt $\rm{J}/\psi$ comes from the fact that their spin state polarization is conceptually and effectively different \cite{NuSea:2000vgl, Faccioli:2022alj}. The measurement of non-prompt $\rm{J}/\psi$ can also provide direct determination of the nuclear modification of beauty hadrons.

In experiments, $\rm{J}/\psi$ is reconstructed through its electromagnetic decay to lepton pairs, in either $e^{+}+e^{-}$ or $\mu^{+}+\mu^{-}$ decay channels. By reconstructing the invariant mass spectra of these lepton pairs ($m_{ee}$ or $m_{\mu\mu}$), one can extract the signal for inclusive $\rm{J}/\psi$ by fitting a suitable signal function and subtracting the background continuum. Usually, a Crystal Ball function~\cite{Gaiser:1982yw} is used as the signal function. To further estimate the non-prompt contribution in the inclusive $\rm{J}/\psi$ signal, one has to rely on the non-prompt production topology. As the beauty hadrons undergo weak decay, the resulting $\rm{J}/\psi$ will originate from a decay vertex that is displaced from the primary interaction vertex (as shown in Fig.~\ref{fig:jpsicartoon}). For this, the pseudoproper decay length ($c\tau$) of the candidate is estimated, which is given in Eq.~\ref{eq:ctau}. The $c\tau$ Probability Density Functions (p.d.f.) for the prompt ($F_{\rm prompt}(c\tau)$) and non-prompt ($F_{\rm B}(c\tau)$) production can be obtained from Monte Carlo simulations separately. By using an unbinned 2-dimensional likelihood fit as described in detail in Refs.~\cite{ALICE:2012vpz,ALICE:2015nvt}, the ratio of the non-prompt to inclusive $\rm{J}/\psi$ production ($f_{\rm B}$) can be estimated, which can be used to calculate the non-prompt and prompt production cross-sections ($\sigma_{\rm{J}/\psi}$), as given below.

\begin{equation}
\begin{split}
\sigma_{{\rm non-prompt}~ \rm{J}/\psi} &= f_{\rm B} \cdot \sigma_{\rm{J}/\psi}, \\
\sigma_{{\rm prompt}~ \rm{J}/\psi} &= (1-f_{\rm B}) \cdot \sigma_{\rm{J}/\psi}
\end{split}
\end{equation}  


Machine learning (ML) techniques are in use in the field of nuclear and particle physics over the last couple of decades~\cite{Denby:1987rk, Lonnblad:1990bi}. Recently, with the advancement of superior hardware and smart algorithms, it has gained its rightful popularity in the big data community. By construction, machine learning is trained to learn the mapping from the input features to the output class. The algorithm helps to learn the correlations between the input and output by optimizing the model parameters on the training data. This is practically useful when the mapping function is not trivial, or sometimes it can not be defined. In such cases, machine learning helps to do the mapping in a faster and more efficient manner, without compromising the quality of the result. The successful application of machine learning techniques in collider experiments is well proven by now. It has been used to tackle many varieties of problems. Some of them include the impact parameter estimation~\cite{Mallick:2021wop,Zhang:2021zxd,Bass:1993vx,David:1994qc,Bass:1996ez}, particle identification and track reconstruction~\cite{Farrell:2018cjr, Belayneh:2019vyx, Goncharov:2021wvd}, jet tagging~\cite{deOliveira:2015xxd, Komiske:2016rsd, Chen:2019uar, Moreno:2019bmu}, anisotropic flow measurements~\cite{Mallick:2022alr, Mallick:2023vgi, Hirvonen:2023lqy},  etc. Interested readers may refer to some of the recent reviews on machine learning in high energy physics~\cite{Albertsson:2018maf, Boehnlein:2021eym, Zhou:2023pti, Feickert:2021ajf}. In this work, for the first time, machine learning techniques are implemented to separate the prompt and non-prompt dimuon pairs from the background to obtain a better identification of the $\rm{J}/\psi$ signal for different production modes. The study has been performed in $pp$ collisions at $\sqrt{s}$ = 7 and 13 TeV simulated using PYTHIA8. Machine learning models such as XGBoost and LightGBM are explored. Some of the motivations of this work are as follows. This technique provides a faster and more efficient method to identify the inclusive, prompt, and non-prompt $\rm{J}/\psi$ signal than the conventional template fitting method discussed above. It can be applied to identify $\rm{J}/\psi$ meson in the entire range of transverse momentum ($p_{\rm T}$) and rapidity ($y$), thus allowing us to probe the production fraction ($f_{\rm B}$) of non-prompt $\rm{J}/\psi$ easily for very fine bins in $p_{\rm T}$ and $y$. This method has another advantage, as it can directly identify the dimuon pairs, hence it can tag them to one of the three sources, prompt, non-prompt, or background. This identification of the dimuon level tags can help in studying many aspects of charmonia and bottomonia production, which are almost impossible using conventional methods. One such application would be the effect of polarization on prompt and non-prompt $\rm{J}/\psi$ production. Apart from these motivations, the novelty of this work also lies in the fact that the attempt to separate prompt versus non-prompt production for $\rm{J}/\psi$ is never attempted before using the machine learning approach.


The paper is organized as follows. It begins with a brief introduction in Sec.~\ref{sec:intro}. The methodology, including the data generation using PYTHIA8, and the description of the machine learning models, are described in Sec.~\ref{sec:method}. The training, evaluation, and quality assurance of the models are discussed in Sec.~\ref{sec:MLtraining} followed by the results and discussions in Sec.~\ref{sec:results}. Finally, the paper concludes by summarizing the findings in Sec.~\ref{sec:summary}.

\section{Methodology}
\label{sec:method}
The descriptions of pQCD-based particle production, such as jets, charm, and bottom hadrons, etc., are well explained by the PYTHIA8 Monte Carlo model. In the current work, we use the PYTHIA8 event generator to simulate the data sets required to train the machine learning model to identify the prompt and non-prompt dimuon signals from the background dimuon pairs. This section provides a brief description of PYTHIA8, along with the different models used in the study.

\subsection{PYTHIA8}
\label{sec:pythia}
\begin{figure}[ht!]
\includegraphics[scale=0.42]{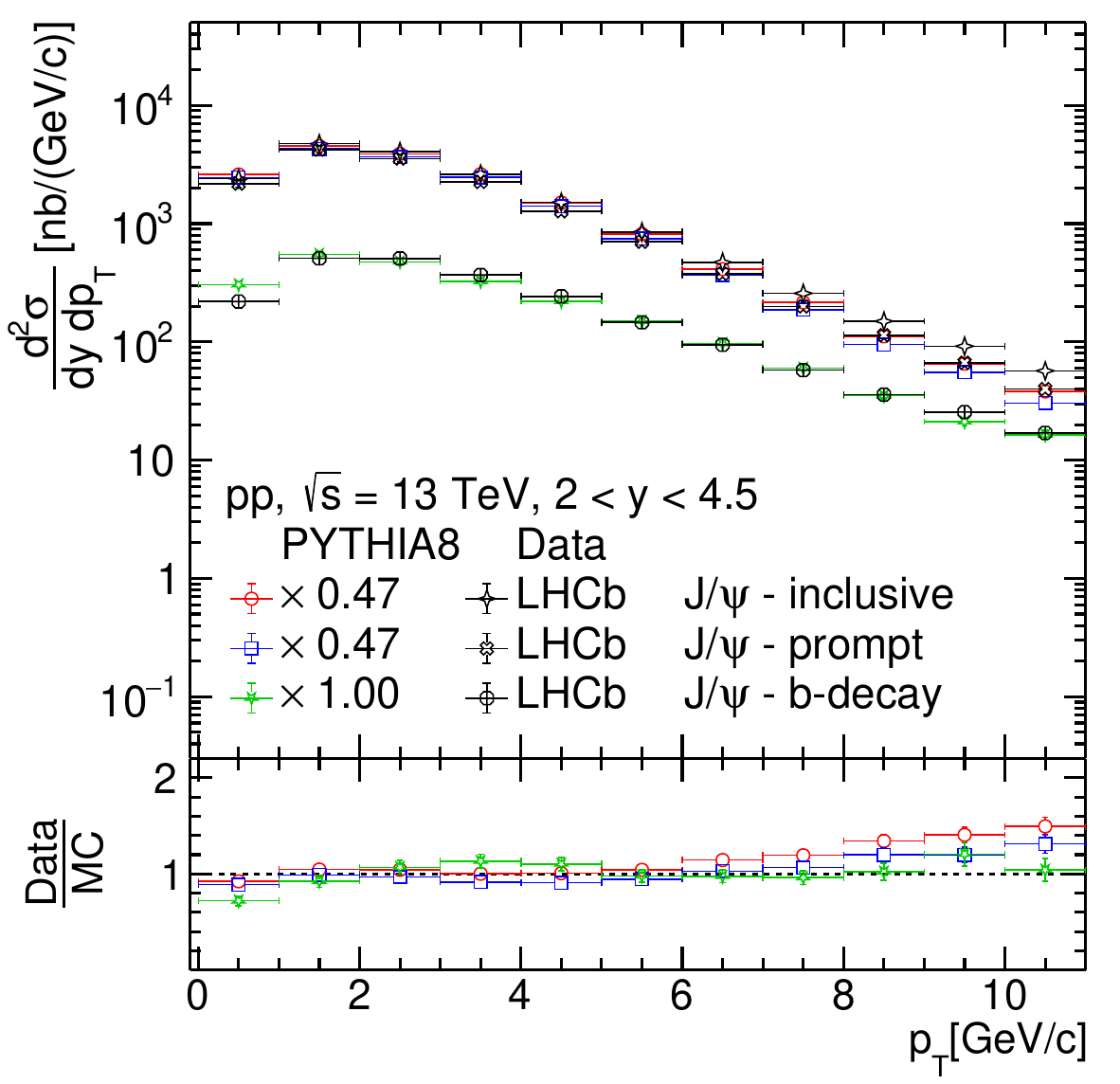}
\caption{Comparision of PYTHIA8 results for inclusive, prompt, and non-prompt production of $\rm{J}/\psi$ meson with the experimental measurements \cite{LHCb:2015foc} in \textit{pp} collisions at $\sqrt{s} = 13$~TeV. A constant multiplication of 0.47, 0.47, and 1.0 is performed to the PYTHIA8 results for inclusive, prompt, and non-prompt production, respectively.}
\label{fig:datacomp}
\end{figure}

PYTHIA is a pQCD-based Monte Carlo event generator used to generate ultra-relativistic \textit{pp} collisions at RHIC and LHC collision energies. PYTHIA8 contains a library of soft and hard processes and models for initial- and final-state parton showers, multiple parton-parton interactions, beam remnants, string fragmentation, and particle decays \cite{Andersson:1983ia, STAR:2010avo}. PYTHIA8 is an improved version of PYTHIA6 where $2\to2$ hard processes are implemented along with MPI-based scenarios to produce the charm and beauty hadrons. In this study, we have used the 4C-tune of PYTHIA8 (see Ref.~\cite{Corke:2010yf} for details) version 8.308 to simulate 20 billion events with inelastic and non-diffractive components (HardQCD:all = on) of the total collision cross section in \textit{pp} collisions at $\sqrt{s} = 13$~TeV and 1 billion minimum bias events in \textit{pp} collisions at $\sqrt{s} = 7$~TeV. The simulation involves a $p_{\rm T}$ cut-off of  $p_{\rm T} > 0.5$ GeV/c (using PhaseSpace:pTHatMinDiverge available in PYTHIA) to avoid the divergence of QCD processes that may occur in the limit $p_{\rm T} \rightarrow 0$. Since this study involves charm and beauty quark production, we have allowed all the charmonia and bottomonia production processes (using ``Charmonium:all=on" and ``Bottomonium:all=on") in PYTHIA8. In addition, we have allowed the spread of the interaction vertex according to a simple Gaussian distribution (Beams:allowVertexSpread=on) where offset and sigma of the spread of the vertices in each of the cartesian axes are taken from Ref.~\cite{ALICE:2010vtz}, and are mentioned in Table~\ref{tab:vertex}. Here, $V_{x}$, $V_{y}$, and $V_{z}$ are the beam interaction vertex distance from the global origin (0,0,0) in the $x$, $y$ and $z$ directions, respectively. We have put an additional cut in the z-vertex, as $|V_{z}|<10$ cm, to be consistent with the experiments. The produced $\rm{J}/\psi$ are allowed to decay in the dimuon channel only, \textit{i.e.} $\rm{J}/\psi\to\mu^{+}+\mu^{-}$ and all other decay modes of $\rm{J}/\psi$ are switched off.


\begin{table}[ht!]
\begin{tabular}{|l|l|l|}
\hline
                      & mean (mm)          & sigma (mm)       \\ \hline 
$V_{x}$            & -0.35           & 0.23        \\ \hline
$V_{y}$              & 1.63           & 0.27        \\ \hline
$V_{z}$            & -4.0           & 40.24         \\ \hline
\end{tabular}
\caption{Offset and sigma values of the primary interaction vertex from the origin.}
\label{tab:vertex} 
\end{table}

Figure \ref{fig:datacomp} shows the comparison of transverse momentum spectra for inclusive, prompt and non-prompt $\rm{J}/\psi$ using PYTHIA8 with the corresponding measurements reported by LHCb \cite{LHCb:2015foc}. All the track cuts for muons and dimuon pairs are kept the same as reported in Ref.~\cite{LHCb:2015foc}.  A factor of 0.47 is multiplied in the PYTHIA8 estimated inclusive and prompt $\rm{J}/\psi$ yields as it overestimates the experimental data. However, PYTHIA8 follows the experimental trend of $p_{\rm T}$ spectra up to $p_{\rm T} < 6\;\rm{GeV/}c$, and starts to deviate towards the higher values of $p_{\rm T}$. One can intuitively note that the yield of $\rm{J}/\psi$ from b-hadron decays is almost ten times lower than the prompt production; however, this difference in production yield between prompt and non-prompt $\rm{J}/\psi$ gets smaller towards the high-$p_{\rm T}$ values. The overall trend produced by PYTHIA8 with the tunes and settings mentioned above is reasonable when compared to the experiment.  The scaling factors are only applied in this plot to match the trend of the experimental data. For all other plots in this work, no such scaling is used and the results are directly from PYTHIA8.

\subsection{Machine learning models}
The realm of ultra-relativistic collisions at the LHC and RHIC produces complex and non-linear systems which demand powerful analysis techniques. These analysis techniques may sometimes require superlative computational facilities yet provide results with significant uncertainties. On the other hand, with the advent of machine learning tools, one can extract insightful results from a vast amount of experimental data with ease and less uncertainty by learning the correlation between the input and target variables. In collider physics experiments, ML models can be exploited in many aspects. One of the complex problems in collider physics experiments is understanding different underlying physical processes that contribute to particle production. However, the final state particles sometimes carry some distinguished kinematic signatures that can help identify their production mechanism and parent particles. For example, in experiments, identifying prompt and non-prompt $\rm{J}/\psi$ meson relies on the statistical separation method, which is already described in Sec.~\ref{sec:intro}. However, using machine learning, one can train a model using some of the kinematic features of the decay daughters to reject the uncorrelated pairs easily and identify the signal and the source of the parent $\rm{J}/\psi$. Popular ML models include gradient-boosting decision trees-based regressions and classifications due to their simplicity, robustness, and efficiency in handling extensive data~\cite{decisionTrees,gbdt}. The name gradient boosting comes because it uses the gradient descent algorithm and boosting method~\cite{gbdt}. In this study, we apply gradient-boosted decision tree-based ML techniques to segregate prompt and non-prompt dimuon pairs from uncorrelated background using the kinematics of all the final state dimuon ($\mu^{+}+\mu^{-}$) pairs, which are discussed below.

\begin{figure}[ht!]
\includegraphics[scale=0.3]{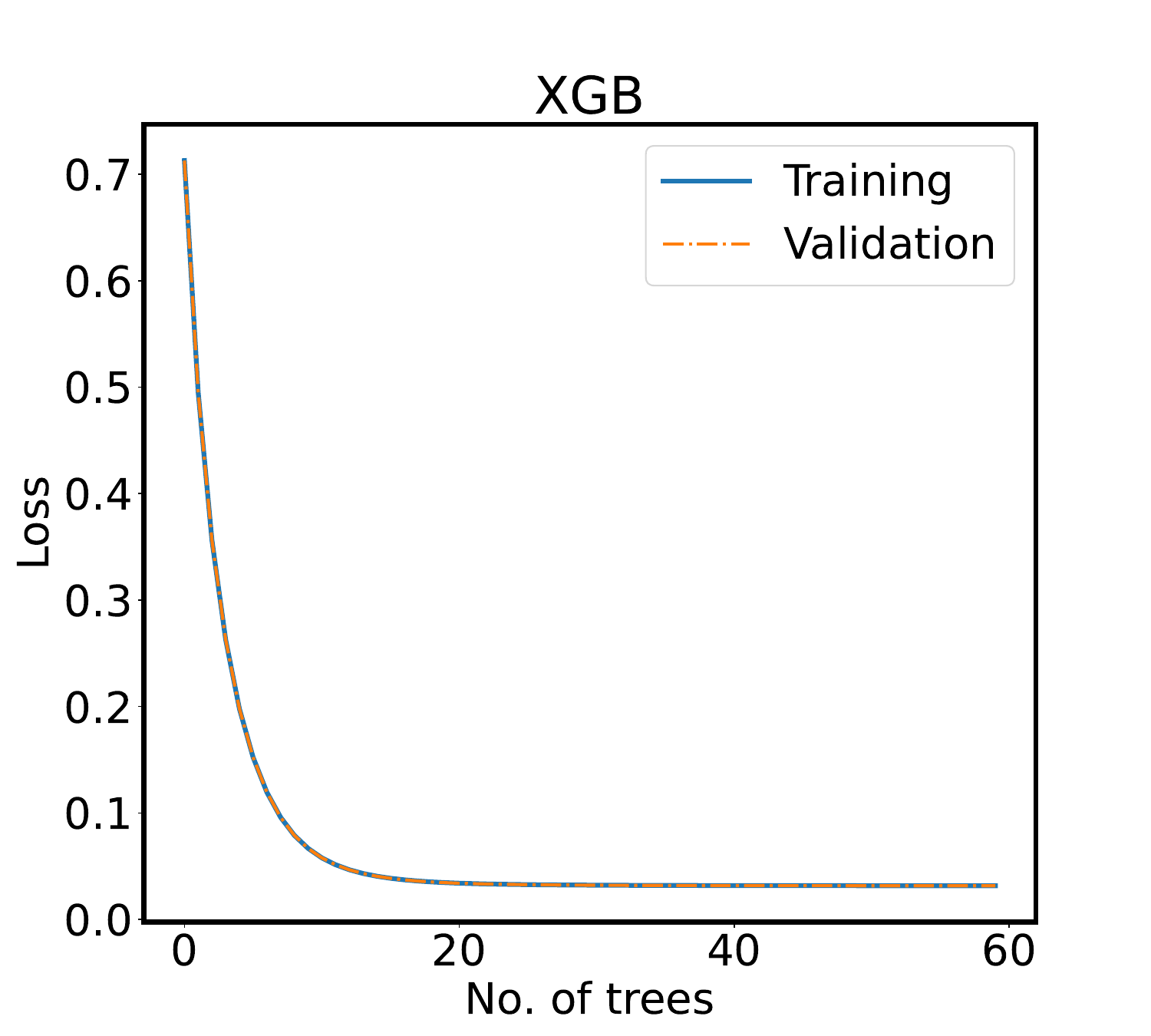}
\includegraphics[scale=0.3]{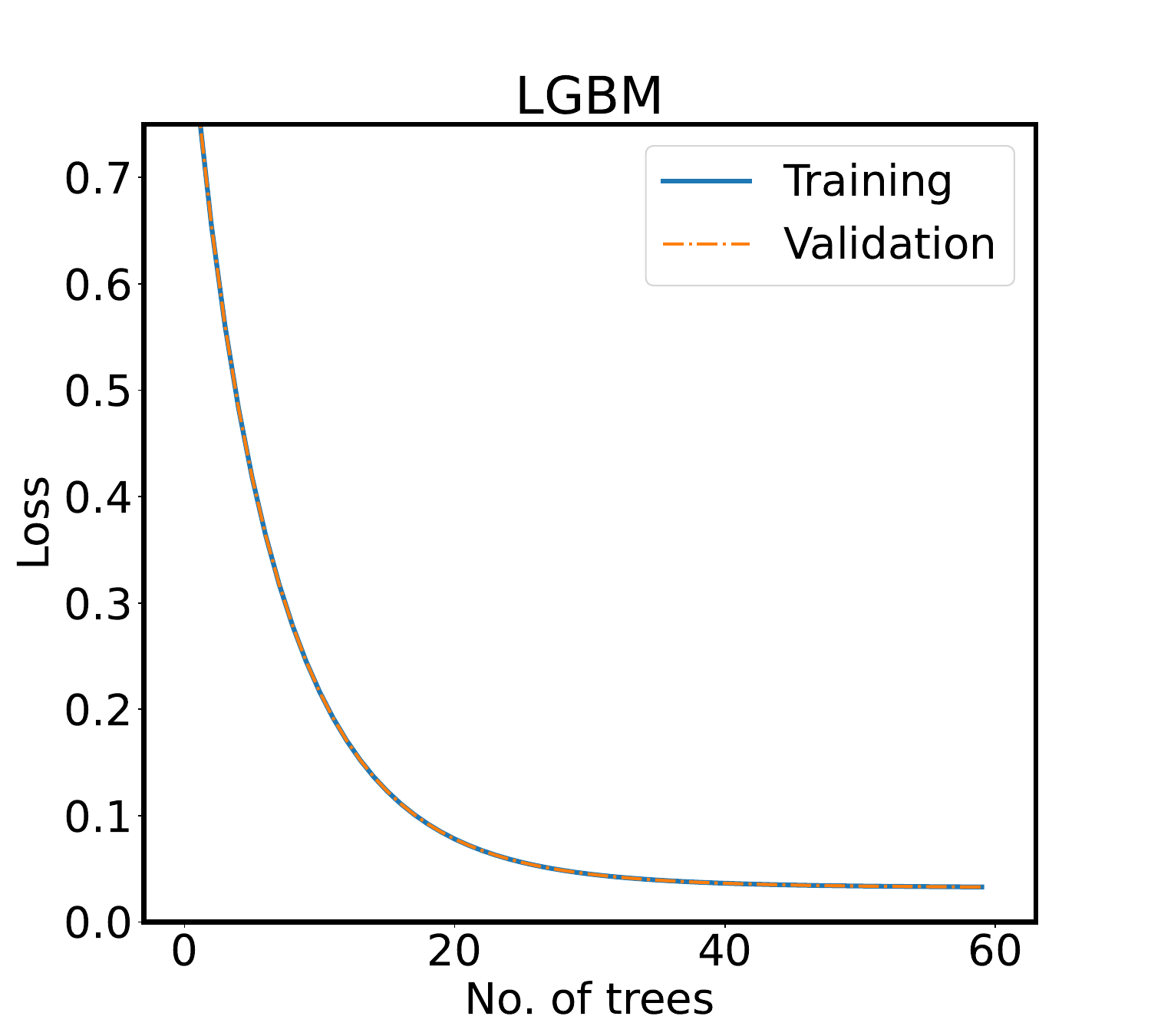}
\caption{ Learning curve (loss versus number of decision trees) for both training (blue) and validation (orange) for both XGB (top) and LGBM (bottom).}
\label{fig:LR}
\end{figure}

\subsubsection{XGBoost}
XGBoost (XGB) \cite{xgboost} stands for Extreme Gradient Boosting, and it is one of the most popular and widely used ML algorithms due to its efficiency in handling large data sets and outstanding performance in classification and regression problems. It is an upgraded version of the gradient-boosting decision trees (GBDT). It has several enhancements, such as parallel computing and tree pruning, to speed up the training process which lets it handle large datasets in a reasonable amount of time. XGB also provides a wide variety of hyperparameters that can be optimized for better model performance~\cite{XGBdocs}.

\subsubsection{LightGBM} 
Light Gradient Boosting Machine (LightGBM or LGBM) \cite{lightgbm} is another enhanced version of the GBDT with improved speed and performance. Along with parallel computing, it uses a leaf-wise splitting of the tree rather than level-wise to increase the model's speed and reduce memory usage. Traditional level-wise splitting of a tree leads to the formation of unnecessary nodes that contain the tiniest information, and these nodes use up memory but do not contribute to the overall learning process. In contrast, splitting a tree leaf-wise leads to the most informative split faster and thus reduces the number of nodes formed, making the training process faster \cite{LGBMdocs}. 

\begin{figure}[ht!]
\includegraphics[scale=0.3]{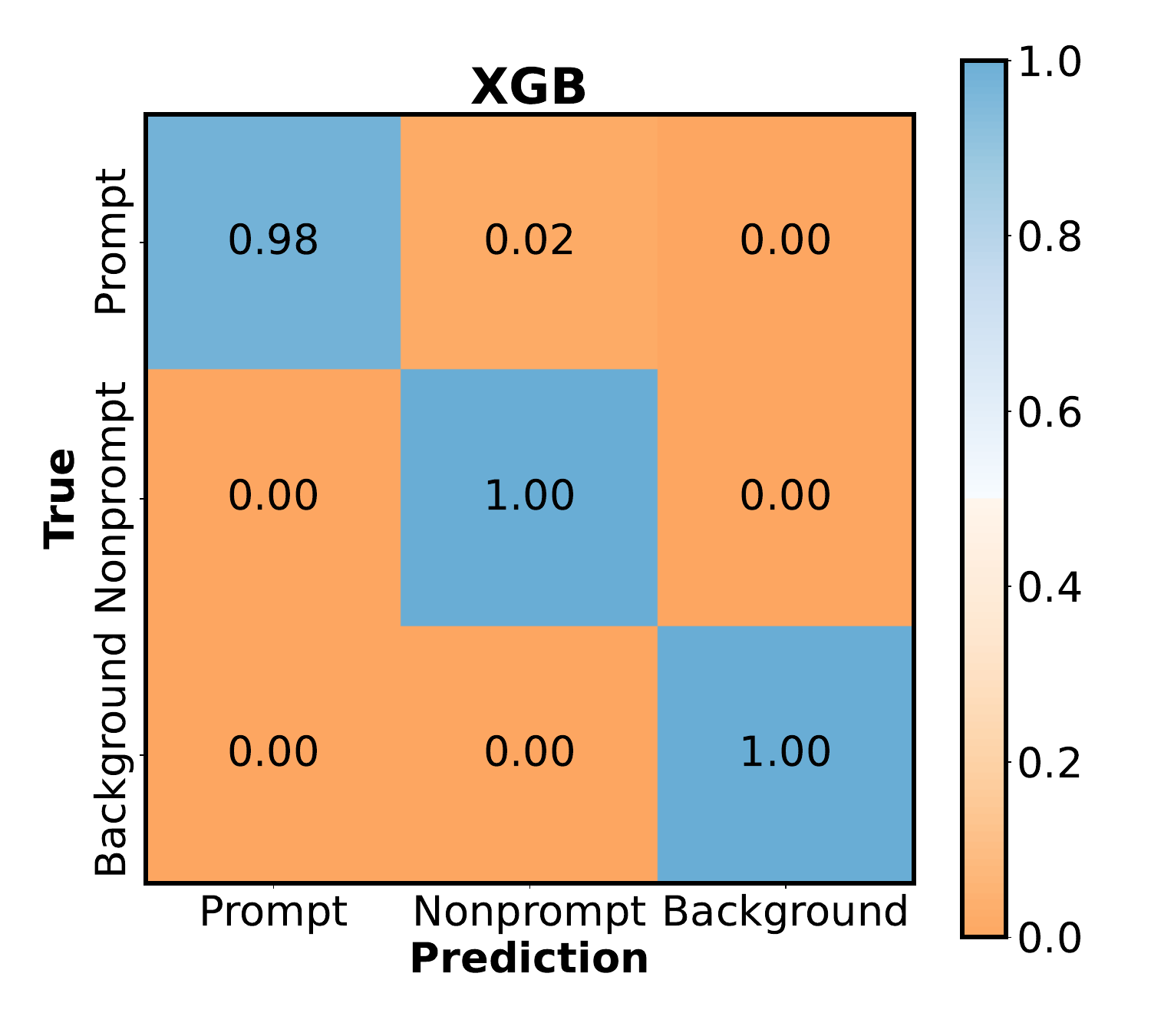}
\includegraphics[scale=0.3]{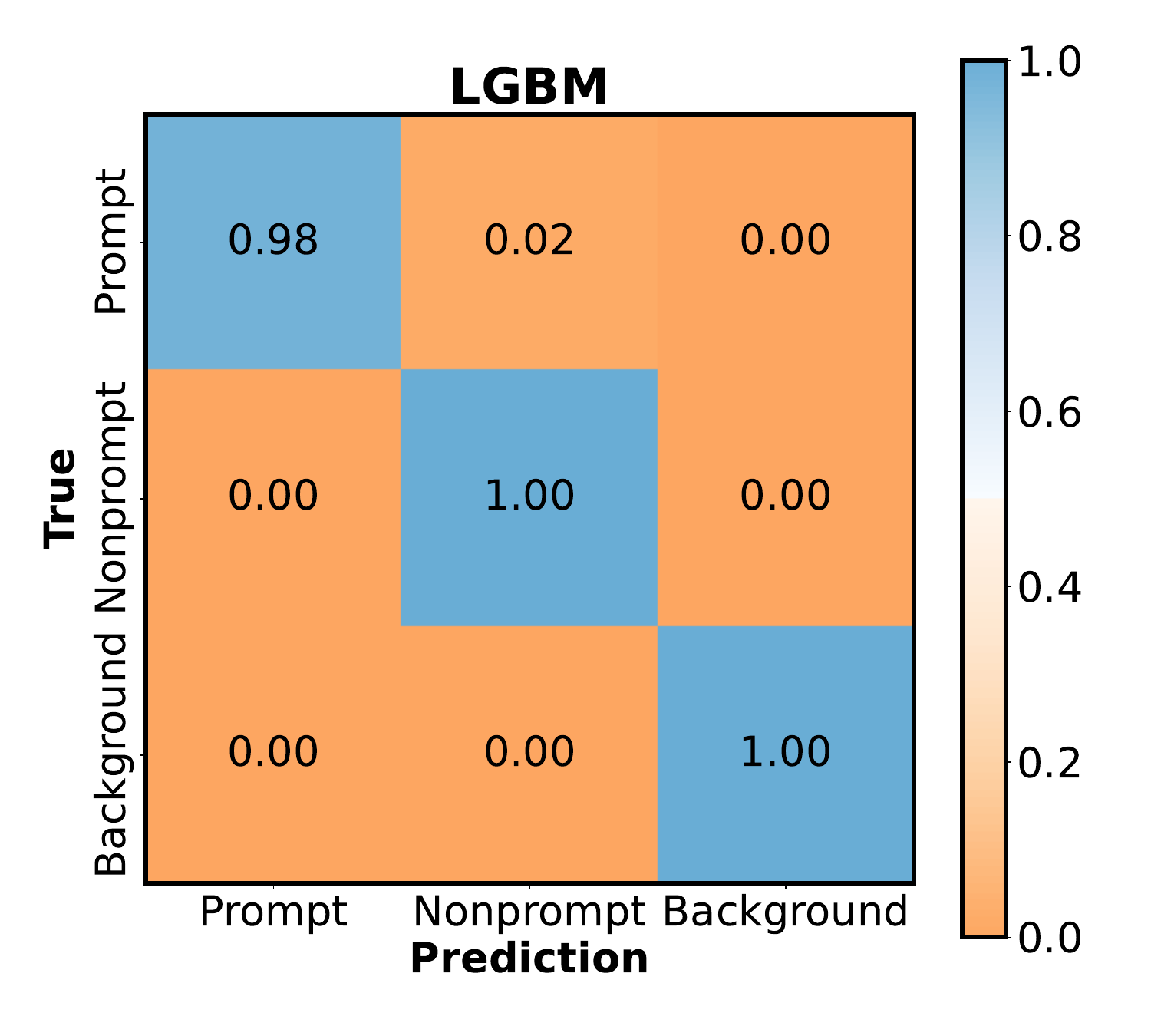}
\caption{ Confusion matrix for both XGB (top) and LGBM (bottom) representing the accuracy and discrepancy in the true and prediction for prompt, non-prompt, and background dimuon pairs.}
\label{fig:CM}
\end{figure}

\section{Training and Evaluation}
\label{sec:MLtraining}

In this section, we discuss our machine-learning models in detail. We begin with the description of the inputs to the models, then preprocessing of the data set, and discuss the model architecture. Finally, we discuss the training and evaluation process with the required quality assurance figures.

\subsection{Input to the machine}
The training of the ML models requires a data set with well-correlated input and target variables. Here, the invariant mass of the reconstructed dimuon pairs ($m_{\mu\mu}$) can significantly help in separating the uncorrelated background from the signal dimuons coming from the $\rm{J}/\psi$ meson. On the other hand, prompt and non-prompt production of $\rm{J}/\psi$ can have different production topologies. The production of the prompt $\rm{J}/\psi$ would be closer to the primary vertex, whereas the $\rm{J}/\psi$ formed from the weak decays of b-hadrons would have a displaced decay vertex with a finite decay length with respect to the primary interaction vertex. One such quantity that is used to differentiate the topological production of the $\rm{J}/\psi$ by taking the production vertex into account is the pseudoproper decay length defined in Eq.~\ref{eq:ctau} below~\cite{ALICE:2021edd}.

\begin{equation}
c\tau=\frac{c\;m_{\rm \rm{J}/\psi}\;\vec{L}\;.\;\vec{p_{\rm T}}}{|\vec{p_{\rm T}}|^{2}}.
\label{eq:ctau}
\end{equation}

Here, $\vec{L}$ is a vector pointing from the primary vertex to the $\rm{J}/\psi$ decay vertex. $c$ is the velocity of light, $m_{\rm \rm{J}/\psi}$ is the mass of $\rm{J}/\psi$ meson taken from the Particle Data Group (PDG)~\cite{ParticleDataGroup:2018ovx}.
For each dimuon pair, we require its invariant mass ($m_{\mu\mu}$), transverse momentum ($p_{\rm{T,\mu\mu}}$), pseudorapidity ($\eta_{\mu\mu}$), and the pseudoproper decay length ($c\tau$) as the input to the models. All these inputs can be obtained in experiments as well. Now, following Eq.~\ref{eq:ctau}, we need the quantity $\vec{L}$ from PYTHIA8, which is obtained using the method described below. One can calculate the $\rm{J}/\psi$ decay vertex for the dimuon pairs using the Eq.~\ref{eq:meetingvertex}.


\begin{equation}
    S_{x}=\frac{(t_1+x_1 m_1/p_{x,1})-(t_2+x_2 m_2/p_{x,2})}{m_1/p_{x,1}-m_2/p_{x,2}}
\label{eq:meetingvertex}
\end{equation}
Here, $S_{x}$ stands for the reconstructed $\rm{J}/\psi$ decay vertex in $x$-direction, for two particles with mass $m_1$ and $m_2$, which fly off from the $\rm{J}/\psi$ decay vertex to a distance $x_1$ and $x_2$, in time $t_1$ and $t_2$ with momentum $p_{x,1}$ and $p_{x,2}$. Similarly, one can also obtain a similar expression for $S_y$ and $S_z$. After obtaining the coordinates for $\rm{J}/\psi$ decay vertex, one can estimate $\vec{L} = \vec{V}-\Vec{S}$. Here, $\vec{V}=(V_{x}, V_{y}, V_{z})$ is the primary vertex coordinates defined in Sec.~\ref{sec:pythia} and $\vec{S}=(S_{x}, S_{y}, S_{z})$ is the $\rm{J}/\psi$ decay vertex position for the reconstructed dimuon pairs, obtained using Eq.~\ref{eq:meetingvertex}.

The target labels for the prompt, non-prompt $\rm{J}/\psi$, and the background dimuon pairs are represented with the numeric tags as 0, 1, and 2, respectively. For the training of the model, the input features are obtained for the opposite sign dimuon pairs in the whole pseudorapidity and transverse momentum range in the minimum bias \textit{pp} collisions at $\sqrt{s} = 13$~TeV using PYTHIA8.

\begin{figure}[ht!]
\includegraphics[scale=0.23]{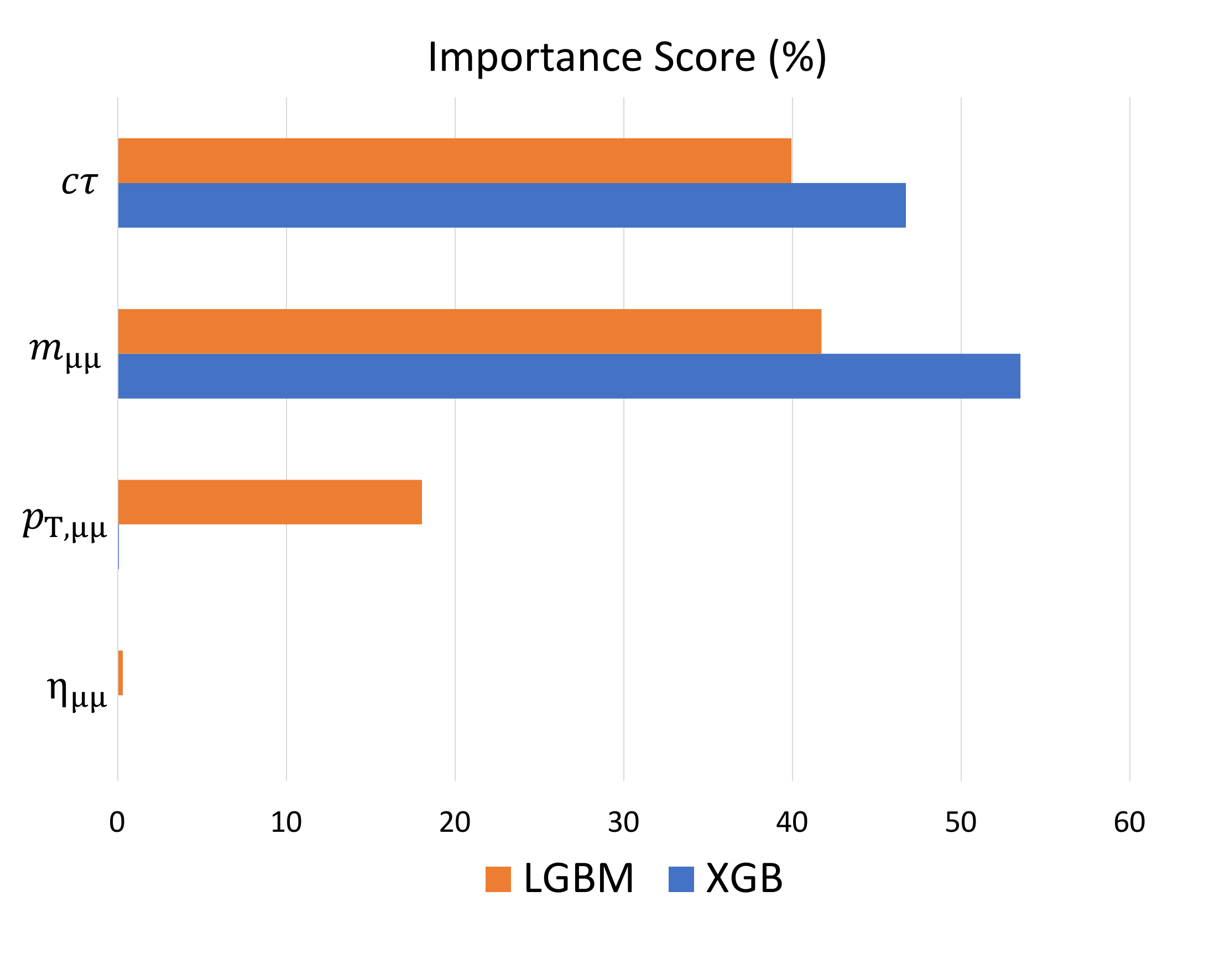}
\caption{ Training importance scores (\%) of pseudoproper decay length ($c\tau$), reconstructed dimuon mass ($m_{\rm \mu\mu}$), transverse momentum ($p_{\rm T,\mu\mu}$) and pseudorapidity ($\eta_{\mu\mu}$) for LGBM (orange), and XGB (blue).}
\label{fig:IS}
\end{figure}

\begin{figure*}[ht!]
\includegraphics[scale=0.27]{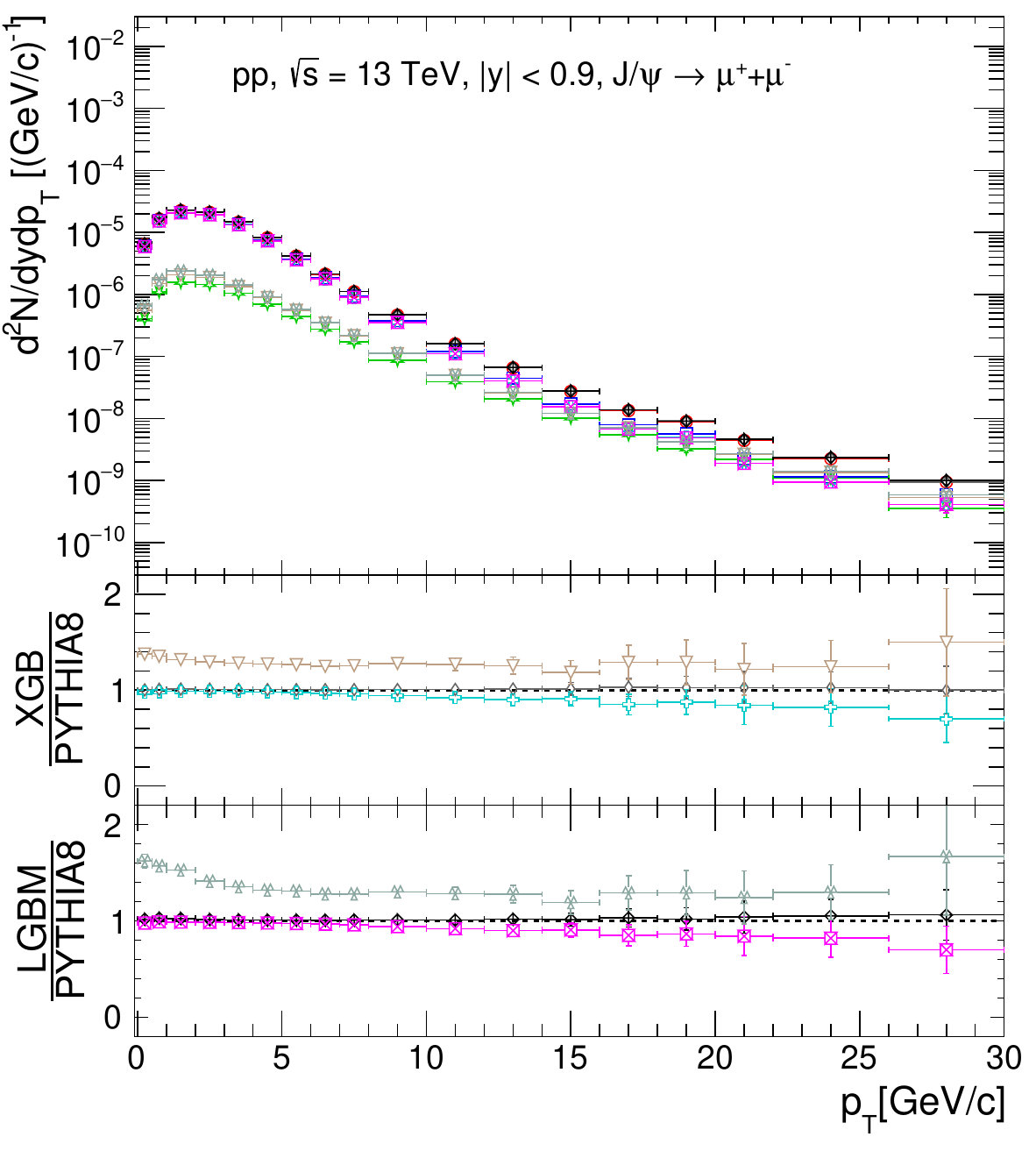}
\includegraphics[scale=0.27]{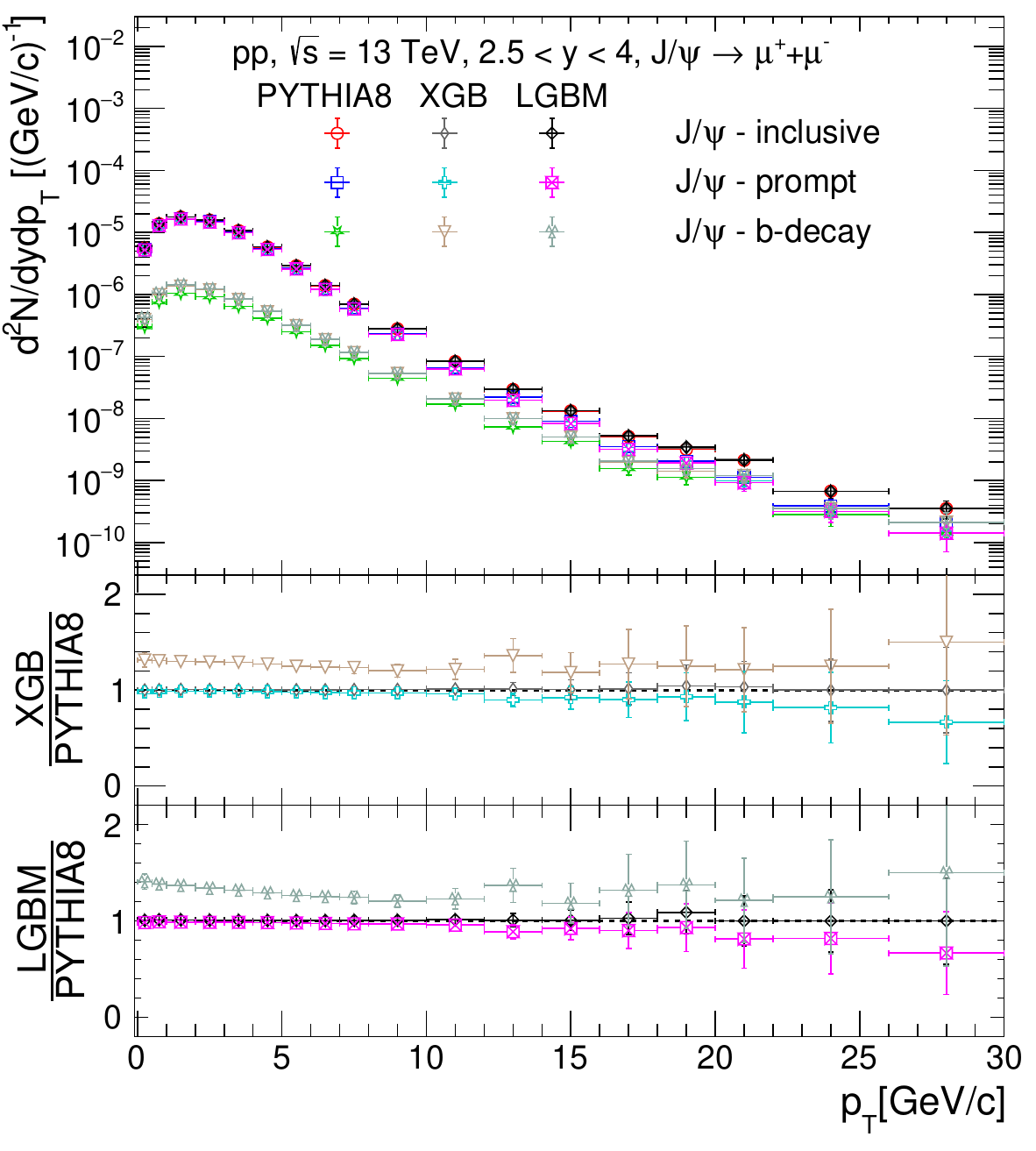}
\includegraphics[scale=0.27]{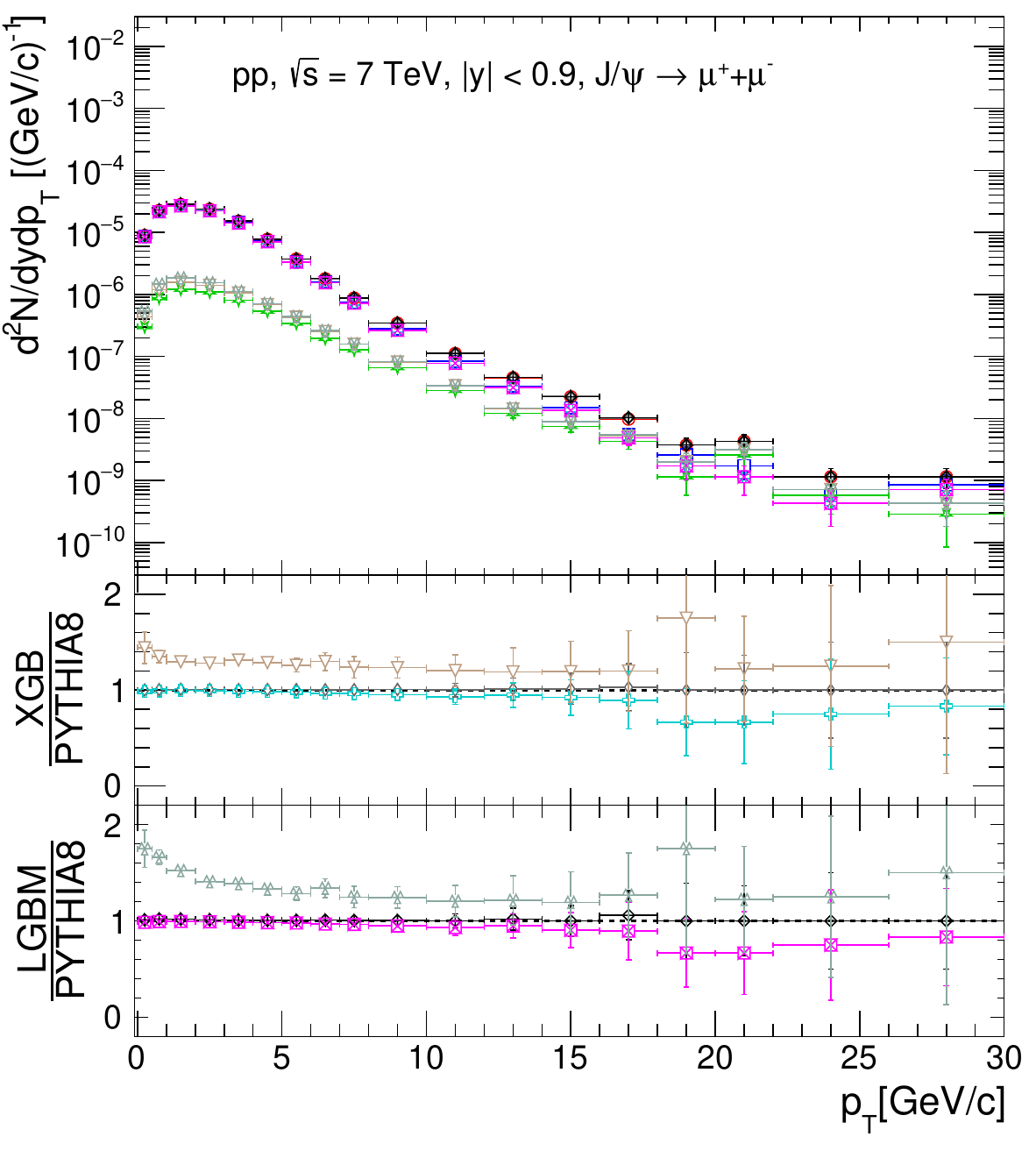}
\caption{ The top panel shows the transverse momentum spectra for the inclusive, prompt and non-prompt J/$\psi$ in \textit{pp} collisions at $\sqrt{s}$ = 13 TeV measured in the midrapidity ($|y|<0.9$) and forward rapidity ($2.5<y<4$), and \textit{pp} collisions at $\sqrt{s}$ = 7 TeV in the midrapidity ($|y|<0.9$) using PYTHIA8 along with the predictions from the XGB and LGBM models. The middle panel shows the ratio of XGB to PYTHIA8, and the bottom panel shows the ratio of LGBM to PYTHIA8.}
\label{fig:ptspectra}
\end{figure*}

\begin{figure}[ht!]
\includegraphics[scale=0.42]{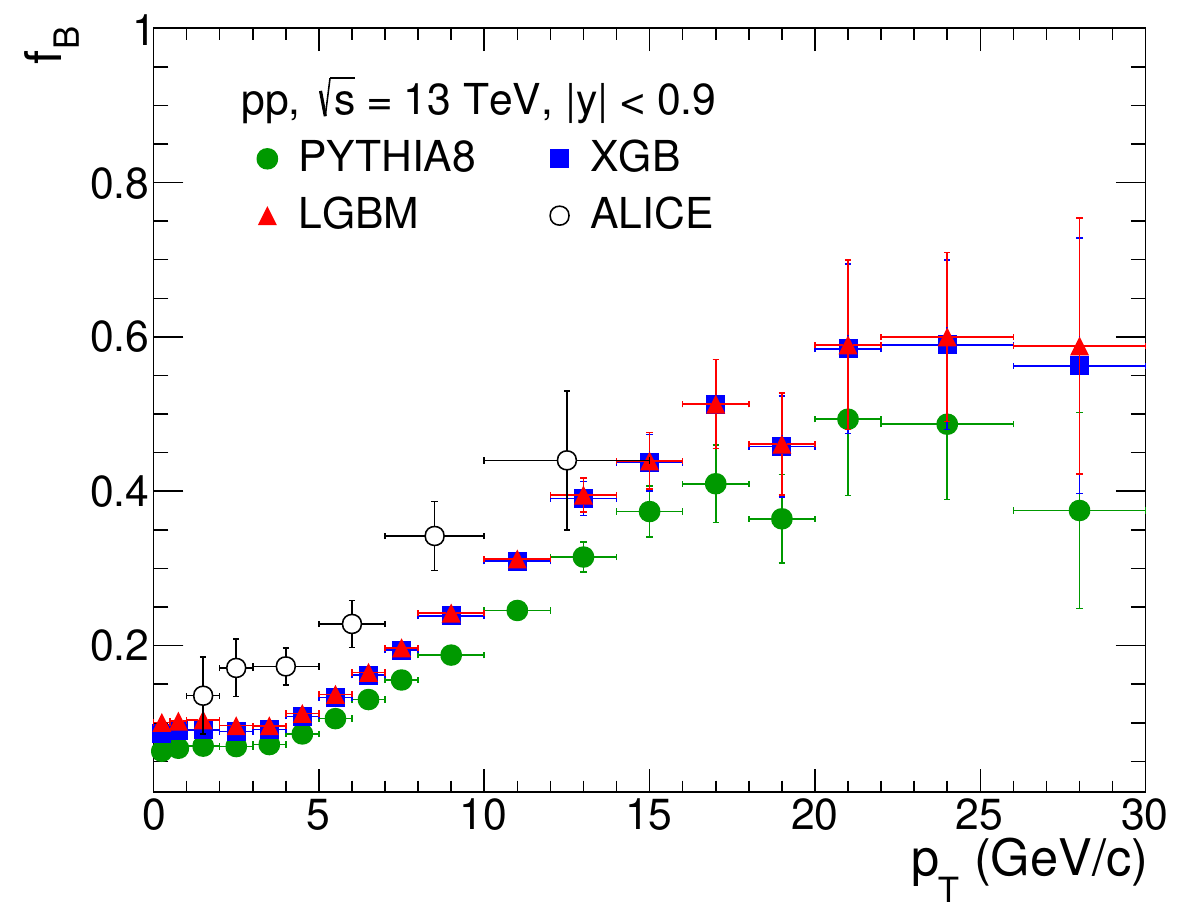}
\caption{Fraction of J/$\psi$ produced from b-hadron decays ($f_{\rm B}$) as a function of transverse momentum at the midrapidity in minimum bias \textit{pp} collisions at $\sqrt{s}$ = 13 TeV using PYTHIA8 and the predictions from XGB and LGBM, compared with the corresponding experimental results from ALICE~\cite{ALICE:2021edd}.}
\label{fig:fB}
\end{figure}

\subsection{Preprocessing and training}

Classification models require to be trained on a similar number of training instances for each of the output classes. We call these instances examples of training. Any imbalance in the examples during the training may bias the output towards the majority class. This is often regarded as the ``class imbalance problem", and the model shows high accuracy just by predicting the majority class. In this study, the majority class is the background followed by the prompt $\rm{J}/\psi$. The ratio of background:prompt:non-prompt is $\approx$ 20:10:1. Thus, the models will favor the training mainly towards the background data, and will mostly misclassify the prompt and the non-prompt $\rm{J}/\psi$. To overcome this data sample imbalance, sampling techniques like undersampling and oversampling are used. Undersampling removes some instances of the majority class while oversampling adds some instances to the minority class to balance the data points present in each class.
Nevertheless, a drawback of undersampling is that it leads to data loss since the instances from the majority class are discarded. Therefore, we prefer to balance the data sets by oversampling. A random oversampling technique from the {\it imblearn} library \cite{Glama} is implemented on the training set wherein both the minority classes (prompt and non-prompt) are resampled to match that of the majority class (background). 
We use 90\% of the entire data as training and the rest 10\% as testing. Further, the resampling is performed on the training set, which solves the class imbalance issue, and then 10\% of the data from the training sample is used as the validation set.

Now, we proceed to define the model architecture and the training process. Model parameters such as the loss function, learning rate, sub-sample, number of trees, and maximum depth are tuned for each model. The best parameters are selected through a grid search method, which is listed in Table \ref{table1}. 

\begin{table}[ht!]
\begin{tabular}{|l|l|l|}
\hline
                      & XGB           & LGBM       \\ \hline 
Learning rate            & 0.3           & 0.1        \\ \hline
Sub-sample              & 1.0           & 1.0        \\ \hline
No. of trees            & 60           & 60         \\ \hline
Maximum depth            & 3             & 3         \\ \hline
Objective      & \textit{softmax} & \textit{softmax} \\ \hline
Metric      & \textit{mlogloss} & \textit{multilogloss} \\ \hline
\end{tabular}
\caption{Parameters used in XGB and LGBM with corresponding values obtained through the grid search method.}
\label{table1} 
\end{table}

In Table~\ref{table1}, the learning rate is a hyperparameter that governs the pace with which the model learns and updates its weights. The subsample indicates the fraction of the data that the model will sample before growing trees, which occurs in every boosting iteration and prevents overfitting. Increasing the maximum depth would make the model more complex. Objective indicates the function that guides the training process, which quantifies the model's performance and reduces the prediction error. In both models, we have used \textit{softmax} objective for the multiclass classification, available as \textit{`multi:softmax'} and \textit{`multiclass'}, for XGB and LGBM, respectively \cite{XGBdocs, LGBMdocs}. The metric is the function that evaluates the model's performance in each training iteration. In both models, we have used the \textit{logloss} metric function for the multiclass classifications, the definition of which can be found in Refs.~\cite{XGBdocs, LGBMdocs}. All the other hyperparameters are kept as their default values for both models.

\subsection{Quality assurance}
\label{sec:qa}

Figure~\ref{fig:LR} shows the learning curve for XGB (top) and LGBM (bottom) for both training and validation, \text{i.e.}, the evolution of the loss as a function of the number of decision trees. For good training, the loss decreases with the increase in the number of decision trees and saturates at a particular loss value, indicating that the training must be stopped now. Another essential training benchmark can be deduced by looking at the difference between the curves for the training and validation simultaneously. For reasonable training, the learning curves for the training and validation should be close; however, a big difference between them can arise due to overfitting or underfitting. One can infer from Fig.~\ref{fig:LR} that the loss values for validation and training decrease with the increase in the number of trees and saturates at around 25 trees for XGB and at around 45 trees for the LGBM. In addition, for both XGB and LGBM, the curves for validation and training lie on top of each other, indicating no overfitting by the models. 

Another essential benchmark of the classification models can be inferred from the confusion matrix or sometimes called as error matrix. Each row of a typical confusion matrix represents the instances of a true class, while each column represents the instances of a predicted class. The confusion matrix as a whole represents the confusion by the model to predict different classes. In Fig.~\ref{fig:CM}, the normalized confusion matrix is shown for XGB and LGBM with the three output classes, \textit{i.e.}, prompt, non-prompt, and the background. Both XGB and LGBM have similar predictions; the backgrounds and the non-prompt dimuon pairs are identified correctly with 100\% accuracy; however, the models misidentify 2\% of the dimuons coming from the prompt $\rm{J}/\psi$ as the non-prompt dimuons. As the ratio of prompt to non-prompt is around 10:1, this discrepancy in the identification has less effect on the prompt; but, it may enhance the non-prompt production yield. Initially, this 2\% misclassification yield coming from the prompt $\rm{J}/\psi$ to the non-prompt $\rm{J}/\psi$, was suspected to be contributed from the indirect prompt production, which are the decays from higher excited states of charmonia. This is because, they might not have produced and decayed exactly at the primary vertex, and therefore may have traveled a finite pseudoproper decay length before decaying. This probable cause is discarded as a similar prediction is obtained while dealing with data set having only indirectly produced prompt $\rm{J}/\psi$. So, this misclassification error is inherited in the model itself.

Figure~\ref{fig:IS} shows the percentage importance score of each feature during training for both XGB and LGBM models. In the context of decision trees, the importance score for a feature is defined as the number of times the feature is used to split a node. The importance score shown in the figure indicates how useful or valuable each feature is during the construction of the boosted decision trees. As one can infer from the figure, the input features that carry the most information about the production species of the reconstructed dimuon pairs are $m_{\mu\mu}$ and $c\tau$, and hence, these are the crucial features for this classification task. In the LGBM model, the order of relative importance to the classification task is $m_{\rm \mu\mu}>c\tau>p_{\rm T,\mu\mu}>\eta_{\mu\mu}$. In contrast, XGB requires only $m_{\rm \mu\mu}$ and $c\tau$ to make a prediction, whereas the model discards the contribution of $p_{\rm T,\mu\mu}$ and $\eta_{\mu\mu}$. Another aspect to learn from this figure is that for the same classification task, different models can learn from the same input features with different importance scores. However, for this classification task,  $m_{\rm \mu\mu}$ and $c\tau$ hold the highest importance scores in both models.

\begin{figure*}[ht!]
\includegraphics[scale=0.35]{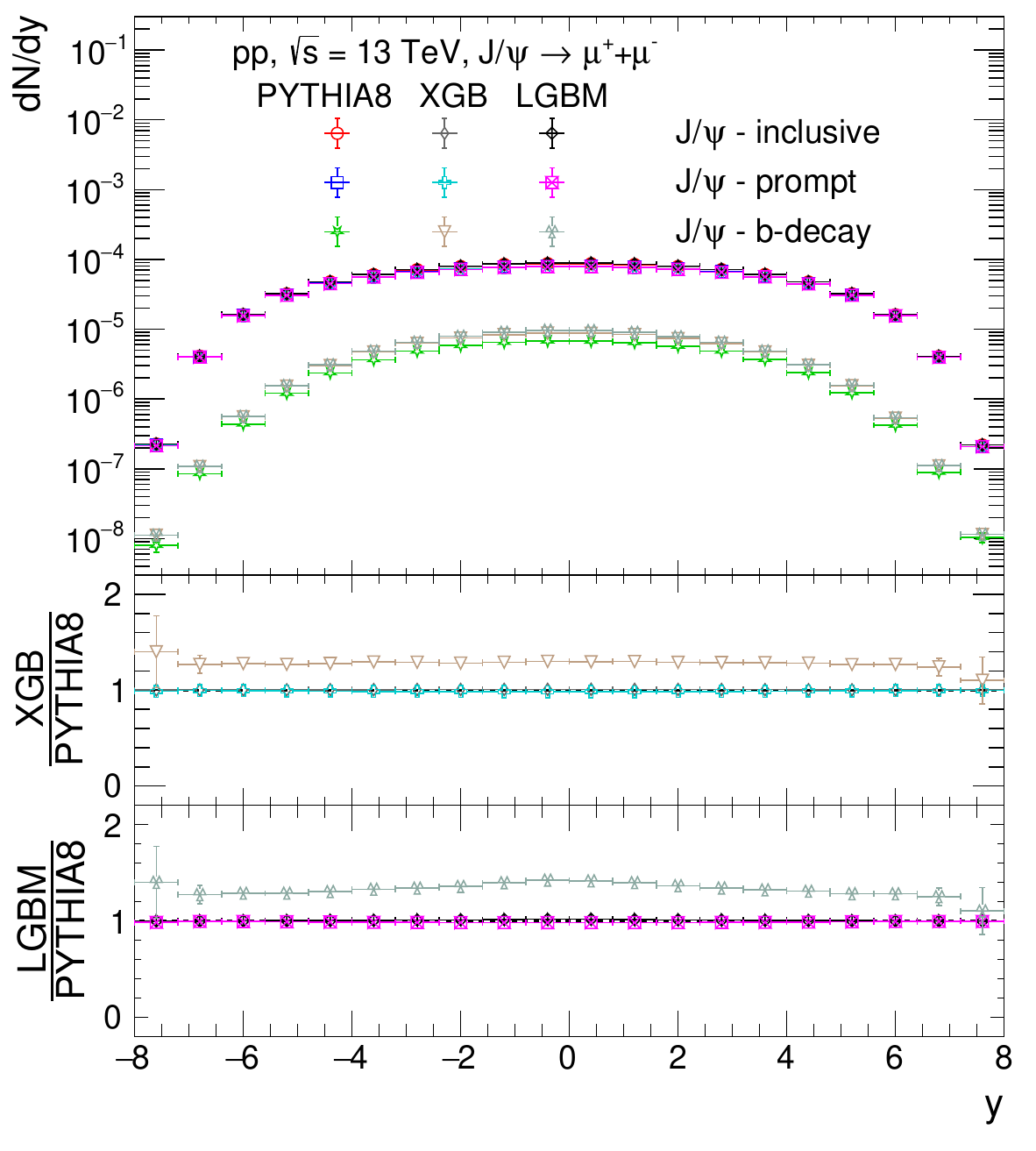}
\includegraphics[scale=0.35]{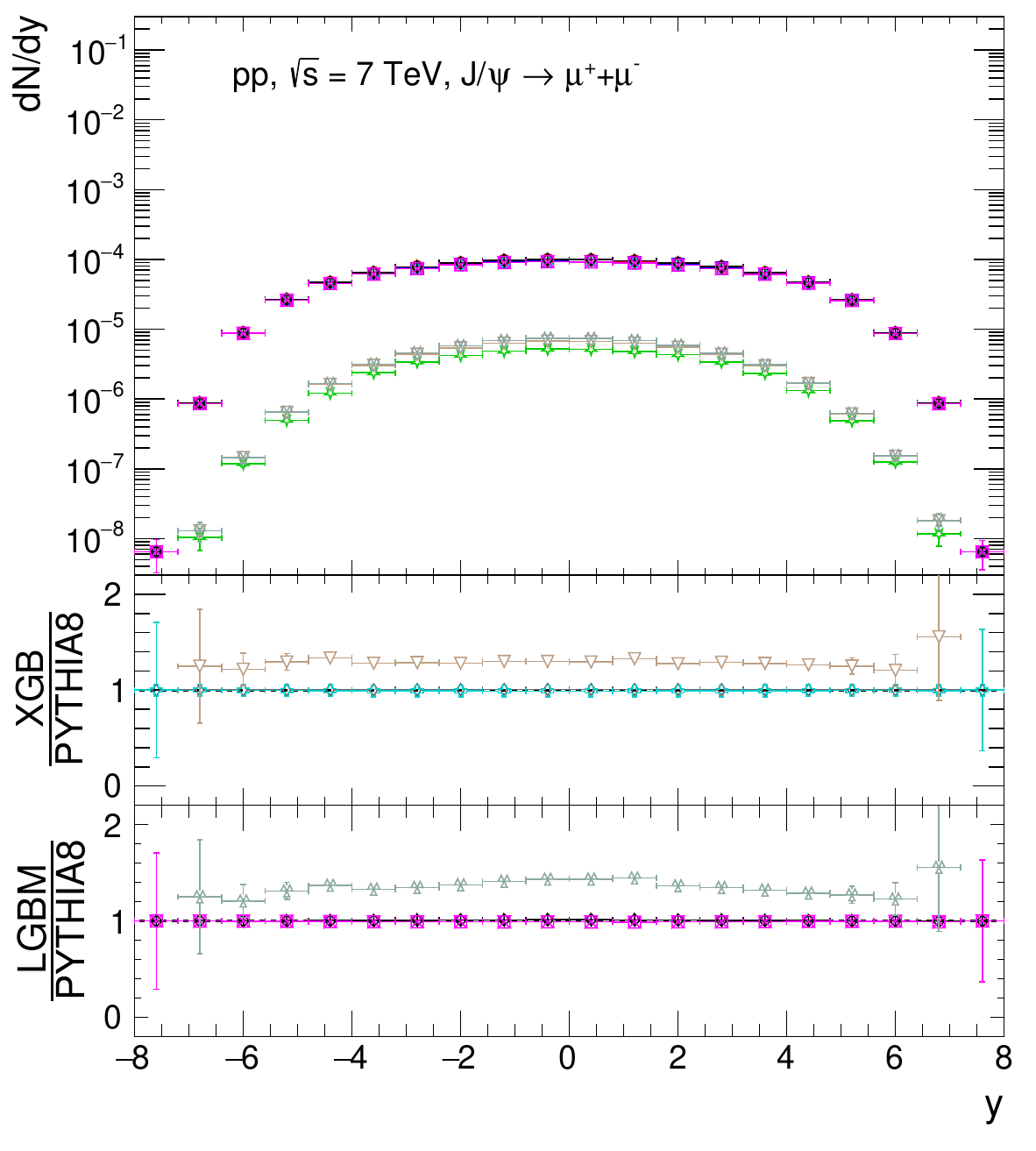}
\caption{Top panel shows the rapidity spectra for inclusive, prompt, and non-prompt production of $\rm{J}/\psi$ in minimum bias \textit{pp} collisions at $\sqrt{s}$ = 13 TeV (left) and $\sqrt{s}$ = 7 TeV (right) using PYTHIA8 and includes the predictions from XGB and LGBM models. The middle panel shows the ratio of XGB to PYTHIA8, and the bottom panel shows the ratio of LGBM to PYTHIA8.}
\label{fig:rapspectra}
\end{figure*}

\section{Results}
\label{sec:results}


Figure~\ref{fig:ptspectra} shows the transverse momentum ($p_{\rm T}$) spectra for the inclusive, prompt and non-prompt $\rm{J}/\psi$ in minimum bias \textit{pp} collisions at $\sqrt{s} = 13$~TeV in midrapidity ($|y|<0.9$) and forward rapidity ($2.5<y<4$). Additionally, the $p_{\rm T}$--spectra for \textit{pp} collisions at $\sqrt{s} = 7$~TeV in midrapidity ($|y|<0.9$) are also added. These results include PYTHIA8 (true), and the predictions from both the trained models \textit{i.e.} XGB and LGBM, which are trained with minimum bias \textit{pp} collisions at $\sqrt{s} = 13$~TeV data. Here, $\rm{J}/\psi\to\mu^{+}+\mu^{-}$ channel is used to reconstruct the $p_{\rm T}$--spectra. At first glance, one notices that the $\rm{J}/\psi$ produced from the b-hadron decays have a significantly lower yield in the low-$p_{\rm T}$ region than the prompt $\rm{J}/\psi$. However, this difference in their production yield tends to decrease as one moves towards high-$p_{\rm T}$. These observations using PYTHIA8 are consistent with the experimental measurements~\cite{ALICE:2021edd, CMS:2010nis, ATLAS:2018hqe}. It is seen that both the machine learning models, XGB and LGBM, can accurately identify the inclusive and prompt dimuon pairs originating from $\rm{J}/\psi$, and thus, their predictions for the $p_{\rm T}$--spectra match well with the results obtained from PYTHIA8 (true). However, some discrepancy arises when both XGB and LGBM models try to identify the dimuon pairs coming from the non-prompt $\rm{J}/\psi$. Both models consistently overestimate the yield of non-prompt $\rm{J}/\psi$. The predictions from the LGBM model are slightly worse at low-$p_{\rm T}$ for the midrapidity case as compared to the XGB model, whereas in the intermediate to high-$p_{\rm T}$, both the models are fairly comparable in accuracy. As discussed earlier in the description of Fig.~\ref{fig:CM}, this overestimation of the yield of the non-prompt $\rm{J}/\psi$ predicted by both the models is a direct consequence of the misidentification of the dimuons coming from the prompt $\rm{J}/\psi$ as the non-prompt dimuons. 

In addition, both XGB and LGBM models are found to be robust for the energy dependence predictions of inclusive, prompt, and non-prompt $\rm{J}/\psi$ $p_{\rm T}$--spectra as seen in Fig.~\ref{fig:ptspectra} for \textit{pp} collisions at $\sqrt{s} = 7$~TeV. It is important to note that the models are trained with $\sqrt{s} = 13$~TeV data, while they can still make predictions for $\sqrt{s} = 7$~TeV. While XGB retains its accuracy of prediction in the entire $p_{\rm T}$ range for the inclusive and prompt $\rm{J}/\psi$ in \textit{pp} collisions at $\sqrt{s} = 7$~TeV, a similar discrepancy for the non-prompt case is observed in \textit{pp} collisions at $\sqrt{s} = 7$~TeV as seen in \textit{pp} collisions at $\sqrt{s} = 13$~TeV. On the other hand, although LGBM retains its accuracy for the inclusive and prompt $\rm{J}/\psi$, it starts to deviate much from the true values towards the lower transverse momentum regions. The success of the models in learning and predicting the energy dependence of inclusive, prompt, and non-prompt production demonstrates the robustness and accuracy of the models. This could be attributed to the fact that most of its learning comes from the invariant mass and the pseudoproper decay length of the dimuon pairs, which are independent of the collision energy.

\begin{figure*}[ht!]
\includegraphics[scale=0.35]{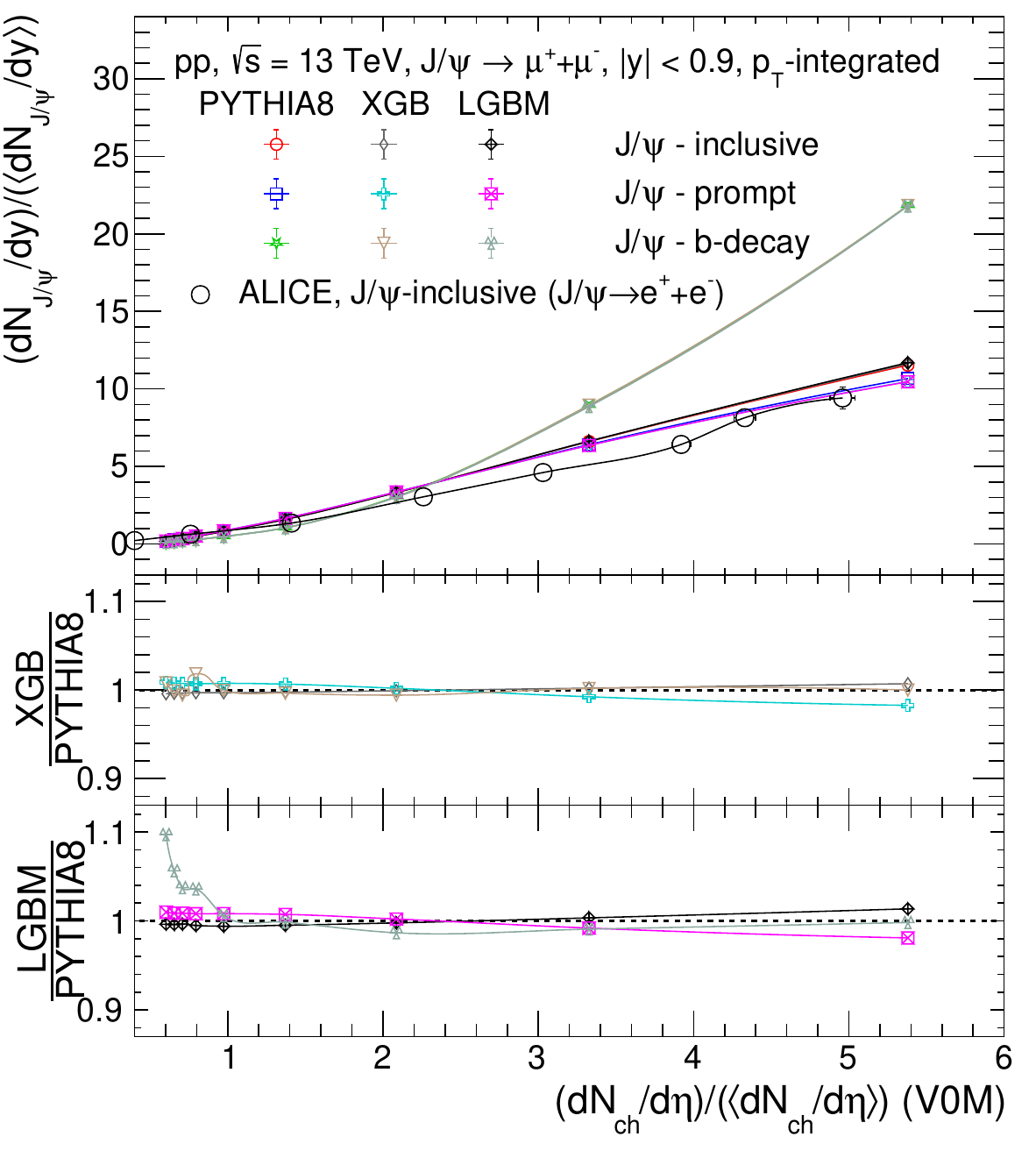}
\includegraphics[scale=0.35]{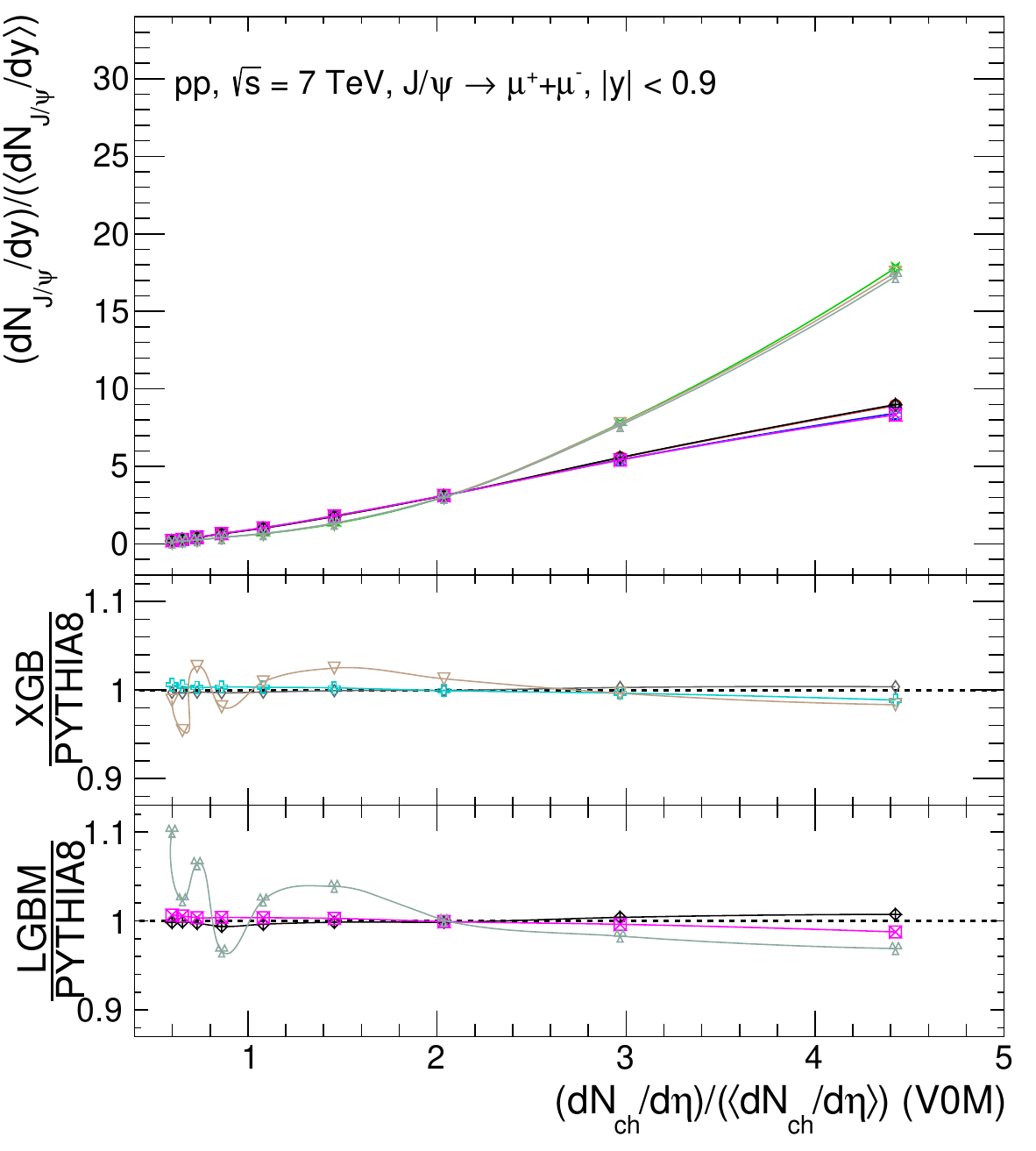}
\caption{Top panel shows the normalized $p_{\rm T}$-integrated inclusive, prompt and non-prompt $\rm{J}/\psi$ yield as a function of normalized charged particle pseudorapidity density at the mid pseudorapidity region with multiplicity selection at the V0 region (V0M) for minimum bias \textit{pp} collisions at $\sqrt{s}$ = 13 TeV (left) and $\sqrt{s}$ = 7 TeV (right) using PYTHIA8 and includes the predictions from XGB and LGBM models, and comparison with experimental data measured at ALICE~\cite{ALICE:2020msa}. The middle panel shows the ratio of XGB to PYTHIA8, and the bottom panel shows the ratio of LGBM to PYTHIA8.}
\label{fig:multiplicity}
\end{figure*}

Figure~\ref{fig:fB} represents the fraction of J/$\psi$ produced from b-hadron decays ($f_{\rm B}$) as a function of transverse momentum at midrapidity in minimum bias \textit{pp} collisions at $\sqrt{s} = 13$~TeV using PYTHIA8. The results are compared with the predictions from XGB and LGBM. The experimental data from ALICE~\cite{ALICE:2021edd} are added. Here, the trend of $f_{\rm B}$ as a function of $p_{\rm T}$ is similar to the experimental observations, where the value of $f_{\rm B}$ is found to be increasing with $p_{\rm T}$ in the range $5.0 \leq p_{\rm T} \leq 20.0$~GeV/c. It is seen that the value of $f_{\rm B}$ remains almost flat and is independent of $p_{\rm T}$ in $p_{\rm T}<5.0$~GeV/c and $p_{\rm T}>20.0$~GeV/c range. By using the machine learning models, we can directly identify the source of the dimuon pairs and hence, it becomes easy to estimate $f_{\rm B}$ in very fine bins of $p_{\rm T}$, that leads to this observation.
As the production fraction of non-prompt $\rm{J}/\psi$ becomes larger in high-$p_{\rm T}$, it is natural to observe that the difference in the $p_{\rm T}$--spectra between prompt and non-prompt $\rm{J}/\psi$ becomes smaller in high-$p_{\rm T}$ as seen in Fig.~\ref{fig:ptspectra}.


Figure~\ref{fig:rapspectra} represents the rapidity spectra for inclusive, prompt, and non-prompt $\rm{J}/\psi$ in minimum bias \textit{pp} collisions at $\sqrt{s}$ = 13 TeV and $\sqrt{s}$ = 7 TeV using PYTHIA8 including the predictions from XGB and LGBM models. The inclusive and prompt $\rm{J}/\psi$ are found to have a flat and rapidity independent yield in the region $|y|<2.5$, after which the yield starts to decrease. On the other hand, for the non-prompt case, the yield is independent of rapidity only for a smaller rapidity coverage, \textit{i.e.}, $|y|<1.0$. These features of the rapidity spectra for different production modes of $\rm{J}/\psi$ using PYTHIA8 are consistent with the experimental measurements reported in Ref.~\cite{ALICE:2021edd}. Interestingly, the predictions from both XGB and LGBM agree with the PYTHIA8 values for the inclusive and prompt $\rm{J}/\psi$ values, while the values for non-prompt $\rm{J}/\psi$ are slightly overestimated. Such a study using a broad range of rapidity is to demonstrate the usefulness and validity of the machine learning models used. However, an experimental
 measurement involving muons is not practical at the mid-rapidity, where the experiment is a multi-purpose one 
 which deals with particle identification, like the ALICE at the LHC and the STAR at the RHIC.

One can observe the magnitude of disagreement in the XGB and LGBM predicted values for the non-prompt $\rm{J}/\psi$ yield with the true values from the simulation is similar to the $p_{\rm T}$--spectra shown in Fig.~\ref{fig:ptspectra} for both the collision energies. For the case of non-prompt $\rm{J}/\psi$, the yield ratio of XGB to PYTHIA8 is almost a constant with a value of 1.3; however, the yield ratio of LGBM to PYTHIA8 is slightly higher in the midrapidity and decreases slowly while moving to forward rapidity. These observations are similar for both collision energies. 


We suspect these discrepancies in the prediction for the non-prompt $\rm{J}/\psi$ are due to the same misidentification of prompt as non-prompt as discussed already in Section~\ref{sec:qa}. However, this discrepancy in the values for the prompt and non-prompt $\rm{J}/\psi$ can be fixed by considering the magnitude of mispredictions in Fig.~\ref{fig:CM}. This is discussed in detail in the Appendix (Sec. \ref{sec:appendix}).

Figure~\ref{fig:multiplicity} depicts the normalized $p_{\rm T}$-integrated $\rm{J}/\psi$ yield for the inclusive, prompt, and non-prompt $\rm{J}/\psi$ as a function of normalized charged particle density at mid-pseudorapidity using PYTHIA8 which includes the predictions from XGB and LGBM models for \textit{pp} collisions at $\sqrt{s} = 13$~TeV and $\sqrt{s} = 7$~TeV. Figure~\ref{fig:multiplicity} also includes the ALICE data comparison for inclusive $\rm{J}/\psi$ yield (measured in the di-electron channel at the mid-rapidity) in \textit{pp} collisions at $\sqrt{s}$ = 13 TeV measured in the V0 region (multiplicity measurement), \textit{i.e.}, $−3.7<\eta<−1.7$ and $2.8<\eta<5.1$~\cite{ALICE:2020msa}. The normalized yields for inclusive, prompt, and non-prompt $\rm{J}/\psi$ from PYTHIA8 are found to increase with the increase in the normalized charged particle density for both the collision energies. The increase in yield is significantly enhanced for the non-prompt $\rm{J}/\psi$, which is consistent with the values reported in Ref. \cite{Weber:2018ddv, Thakur:2017kpv}. While PYTHIA8 slightly overestimates the experimental data, it almost maintains the overall trend of the normalized yield for the inclusive $\rm{J}/\psi$.  Towards higher multiplicities in the final state, $\rm{J}/\psi$ from b-decays show an increasing trend
with non-linear behavior. The slopes of these multiplicity-dependent yields of inclusive, prompt and non-prompt 
$\rm{J}/\psi$ show energy dependence with higher slopes at higher collision energies.
The predictions from XGB and LGBM give an overall good estimation for PYTHIA8 while deviating around 10\% towards the lower multiplicity for the non-prompt  $\rm{J}/\psi$ cases for both collision energies.



\section{Summary}
\label{sec:summary}

In this work, an effort is made to disentangle the inclusive, prompt, and non-prompt $\rm{J}/\psi$ from the uncorrelated background dimuon pairs using machine learning tools. 
We use experimentally available inputs for the models. The $\rm{J}/\psi$ meson are reconstructed in the $\mu^+ + \mu^-$ decay channel. For each dimuon pair, we require its invariant mass ($m_{\mu\mu}$), transverse momentum ($p_{\rm{T,\mu\mu}}$), pseudorapidity ($\eta_{\mu\mu}$), and the pseudoproper decay length ($c\tau$) as the input to the models. We use XGBoost and LightGBM models for this classification task. The training of the models is performed with the minimum bias \textit{pp} collisions at $\sqrt{s} = 13$~TeV simulated with PYTHIA8. The predictions from both the models are tested for \textit{pp} collisions at $\sqrt{s} = 13$~TeV and \textit{pp} collisions at $\sqrt{s} = 7$~TeV. Both the models show accuracy up to 98\%; however, they mis-identify 2\% of the prompt J/$\psi$ as the non-prompt. The transverse momentum ($p_{\rm T}$) and pseudorapidity ($\eta$) differential measurements of inclusive, prompt, and non-prompt $\rm{J}/\psi$, its multiplicity dependence, and the $p_{\rm T}$ dependence of fraction of non-prompt ($f_{\rm B}$) are shown. These results are compared to experimental findings wherever possible. 

This study presents a unique method to separate the production of prompt and non-prompt $\rm{J}/\psi$ from the uncorrelated background dimuon pairs. As the models do not include any fitting to the $p_{\rm T}$ differential spectra, it can be applied to identify each dimuon pairs separately having any value of $p_{\rm T}$ in any rapidity range and thus allow us to probe the production fraction, $f_{\rm B}$ of non-prompt $\rm{J}/\psi$ even in fine bins of $p_{\rm T}$, $\eta$ and $y$. The direct identification of dimuon pairs as prompt or non-prompt can help study many aspects of charmonia and bottomonia production, which are almost impossible using conventional methods. One such application would be the effect of polarization
on prompt and non-prompt $\rm{J}/\psi$ production. 

In addition, ALICE has reported the non-linearity in the normalized $\rm{J}/\psi$ yield at the midrapidity in the dielectron channel towards higher final state normalized multiplicity \cite{ALICE:2021zkd}. As seen in this present study, such behavior is an outcome of the non-prompt 
$\rm{J}/\psi$ both at the mid- and forward rapidities. The present method can be used in the experiments to separate prompt from non-prompt $\rm{J}/\psi$ and hence study the related production dynamics.

\section*{Acknowledgements}

 S.P. acknowledges the financial support from UGC, the Government of India. The authors sincerely acknowledge the DAE-DST, Government of India funding under the Mega-Science Project – “Indian participation in the ALICE experiment at CERN” bearing Project No. SR/MF/PS-02/2021-IITI (E-37123). The authors gratefully acknowledge the usage of resources of the LHC grid Tier-3 computing facility at IIT Indore. The authors would like to thank Mr. Preet Bhanjan Pati, Master's student from IISER Tirupati for the initial exploration
 of this work, the preliminaries of which formed his master's thesis.

\section{Appendix: Bin-wise Yield Correction}
\label{sec:appendix}

\begin{figure*}[ht!]
\includegraphics[scale=0.35]{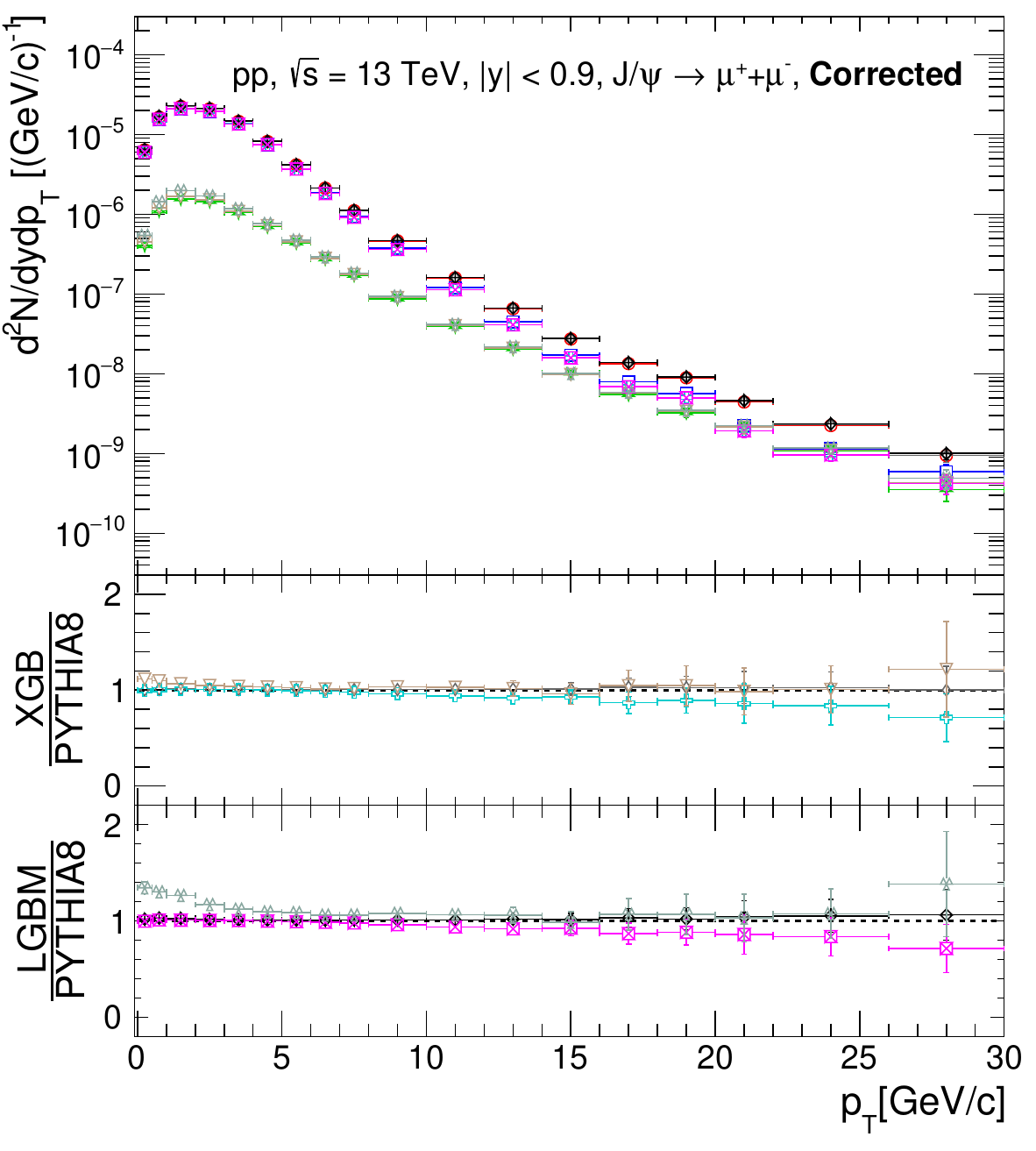}
\includegraphics[scale=0.35]{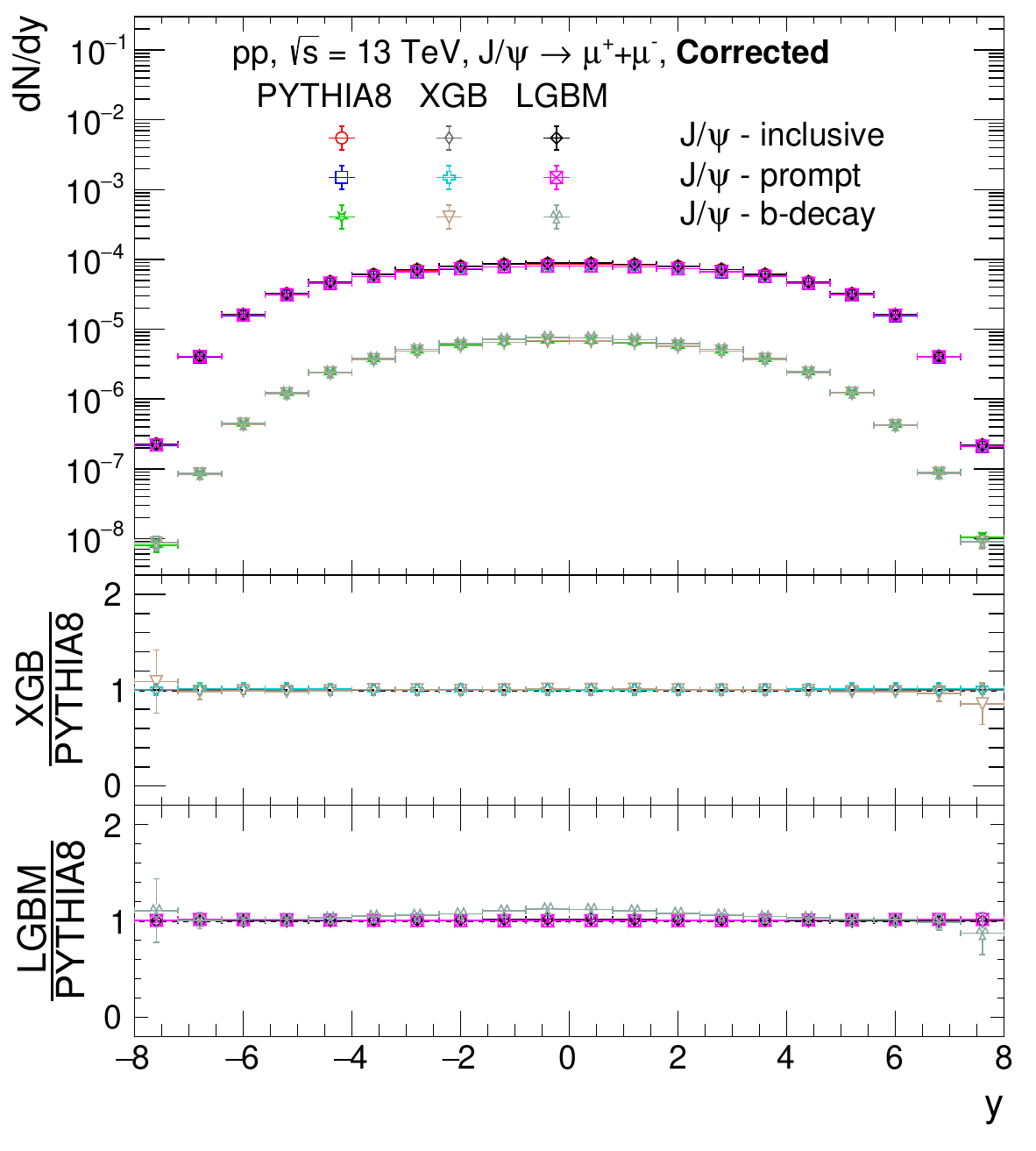}
\caption{Transverse momentum spectra (left), rapidity spectra (right) in minimum bias \textit{pp} collisions at $\sqrt{s}$ = 13 TeV for inclusive, prompt and non-prompt J/$\psi$ using PYTHIA8 compared with the predictions from XGB and LGBM with corrections.}
\label{fig:corrected}
\end{figure*}

The inconsistency between the true and the predicted values for the non-prompt $\rm{J}/\psi$, as shown in Figs.~\ref{fig:ptspectra} and \ref{fig:rapspectra} can be corrected by correcting the yields of each bin in the transverse momentum, rapidity and spectra. Considering the correction factor is independent of transverse momentum, rapidity, and pseudo-rapidity, for a given bin $i$, the corresponding corrected prompt and non-prompt $\rm{J}/\psi$ yield is given by the following expression:
\begin{equation} 
Y_{\rm p, i}^{\rm corr}=\frac{Y_{\rm p, i}^{\rm uncorr}}{1-\rm{f}}
\label{eq:corrprompt}
\end{equation}

\begin{equation}
    Y_{\rm np, i}^{\rm corr}= Y_{\rm np, i}^{\rm uncorr}-\frac{\rm{f}}{1-\rm{f}}\frac{Y_{\rm np, i}^{\rm uncorr}~Y_{\rm p}^{\rm uncorr}}{Y_{\rm np}^{\rm uncorr}}
\label{eq:corrnonprompt}
\end{equation}

In Equations \ref{eq:corrprompt} and \ref{eq:corrnonprompt}, $\rm f$ denotes the correction factor for the prompt yield, which is 0.02 in our case for both XGB and LGBM. $Y_{\rm p, i}$ and $Y_{\rm np, i}$ are the yields in the $i$th bin of the spectra for the prompt and non-prompt, respectively, whereas $Y_{\rm p}$ and $Y_{\rm np}$ denotes the total yields. The superscripts `$\rm{corr}$' and `$\rm{uncorr}$' stand for the corrected and uncorrected yields, respectively. Figure \ref{fig:corrected} shows the corrected transverse momentum and rapidity spectra in minimum bias \textit{pp} collisions at $\sqrt{s}$ = 13 TeV for the prompt, non-prompt and the inclusive $\rm{J}/\psi$. The transverse momentum spectra are calculated in the midrapidity ($|y| < 0.9$) regions. As one can see, the non-prompt $\rm{J}/\psi$ predictions after implementing the corrections using Eq.~\ref{eq:corrnonprompt} matches the PYTHIA8 calculations quite well for all the spectra. The reason for not making such a correction in the present study is to demonstrate the actual predictions given by the ML models.

\end{document}